\pdfoutput=1

\documentclass[amsmath,superscriptaddress,citeautoscript,twocolumn,
showpacs,floatfix,prb]{revtex4}
\usepackage{amssymb,amsmath}
\usepackage{graphics}
\usepackage{epstopdf}
\usepackage{epsfig}
\usepackage{color}
\usepackage{amsfonts}
\usepackage{bm}
\usepackage{amscd}
\makeatother

\begin{document}

\title{Switchable valley functionalities of an $n-n^{-}-n$ junction in 2D semiconductors}

\author{Matisse Wei-Yuan Tu}

\affiliation{Department of Physics and Center of Theoretical and Computational
Physics, University of Hong Kong, Hong Kong, China}

\author{Wang Yao}
\email{wangyao@hku.hk}

\affiliation{Department of Physics and Center of Theoretical and Computational
Physics, University of Hong Kong, Hong Kong, China}
\begin{abstract}
We show that an $n-n^{-}-n$ junction in 2D semiconductors can flexibly realize two basic valleytronic functions, i.e. valley filter and valley source, with gate controlled switchability between the two. Upon carrier flux passing through the junction, the valley filter and valley source functions are enabled respectively by intra- and inter-valley scatterings, and the two functions dominate respectively at small and large band-offset between the $n$ and $n^{-}$ regions.
It can be generally shown that, the valley filter effect has an angular dependent polarity and vanishes under angular integration, by the same constraint from time-reversal symmetry that leads to its absence in one-dimension. These findings are demonstrated for monolayer transition metal dichalcogenides and graphene using tight-binding calculations. We further show that junction along chiral directions can concentrate the valley pump in an angular interval largely separated from the bias direction, allowing efficient havest of valley polarization in a cross-bar device.
\end{abstract}
\maketitle

\section{Introduction}

Exploring diversified internal quantum degrees of freedom of carriers as the fundamental ingredients for device applications has led to the emergence of spintronics and pseudospintronics,
attempting to go beyond present-day charge-based electronics. Utilizing the valley degree of freedom, an inherited property of a plethora of crystalline materials, for building device components has conceived the field of valleytronics.\cite{Gunawan06155436,Gunawan0686404,Rycerz07172,Xiao07236809,Yao08235406} Two-dimensional (2D) hexagonal materials such as graphene and transition metal dichalcogenides (TMD) are especially attractive for valleytronics,\cite{Xu14343,Schaibley1616055} with a time-reversal pair of valleys well separated in momentum space.

A widely adopted approach for producing valley currents in these 2D crystals is to introduce lateral junctions as scatterers.
Novel uses of lateral junctions have been discussed in graphene exploiting the nature of massless Dirac fermions,
for example, the effects of Klein tunneling\cite{Katsnelson06620,Wilmart1401106}, electronic analogs of Veselago's lens\cite{Cheianov071252,Lee15925}, and guiding\cite{Williams11222} at bipolar or unipolar junctions. Valleytronic functionalities arise from the valley contrasted scattering by the lateral junctions, for which various forms of realization have been proposed such as gated regions,\cite{Garcia-Pomar08236801,Pereira09045301,Chen16075407} strained areas,\cite{Zhai10115442,Wu11176802,Jiang13046601,Grujic14046601} or line defects.\cite{Gunlycke11136806,Lv12045410,Liu13195445,Chen14121407,Pulkin16041419,Cheng16103024}
Most of these approaches use line scatters with momentum conserved in the parallel direction, such that the scattering is effectively one-dimensional (1D).\cite{Garcia-Pomar08236801,Pereira09045301,Zhai10115442,Jiang13046601,Wu11176802,Grujic14046601,Chen16075407,
Gunlycke11136806,Liu13195445}
The key mechnism exploited is the valley-dependent intra-valley scattering at oblique incidences upon the line scatters.\cite{Garcia-Pomar08236801,Pereira09045301,Chen16075407,Zhai10115442,Wu11176802,Jiang13046601,Grujic14046601,Gunlycke11136806} In such case, the lateral junction induces valley currents by permeating carriers of one valley more than the other, without producing net outward valley-polarized flows. This defines what valley filters are, as depicted by the left panel of Fig.~\ref{schematics}(a).

It is worth noting that such valley filtering effect does not occur in intrinsic 1D systems (e.g. nanoribbons), where time-reversal symmetry dictates equal amounts of intra-valley reflections for incidences from either valleys.\cite{arXiv:1602.06633}
For the effective 1D scattering event projected from 2D systems with line scatters, the symmetry between incidences at the two valleys can be effectively broken by the oblique incident angle.
The magnitude and polarity of the valley filtering therefore depend on the incident angle.
Concerning the absence of valley filtering in intrinsic 1D systems, the following question naturally arises. In the 2D geometries, can the valley filtering effect still give rise to nonzero valley flux under the integration over incident angles?
In experiments for studying angle-dependent charge transport, the angle dependence is approached by pre-fixing the orientations of local contacts.\cite{Lee15925,Williams11222,Sutar124460,Rahman15013112} With such inflexibility of angular probing and possible diffusive motions that blur the angular dependence, angle-integrated valley fluxes are practically preferable.

On the other hand, lateral junctions also induce inter-valley scattering, which can be non-negligible in addressing the valley functionalities.\cite{Chen16075407,Lv12045410,Cheng16103024,arXiv:1602.06633} Interestingly, it is recently noted that inter-valley scattering can provide a useful resource for valleytronics, besides its well anticipated role in depolarizing valley.\cite{arXiv:1602.06633} In 1D systems, it is shown that inter-valley scattering by disorders can realize a distinct valleytronic functionality, the valley source, where upon passing charge current, valley currents are pumped in both the forward and backward directions, with a net outward valley flux (c.f. the right panel of Fig.~\ref{schematics}(a)).\cite{arXiv:1602.06633} The exploration of such effect in the more relevant 2D scattering geometry is of importance for practical implementation based on 2D crystals.



Here we show that in 2D scattering by line scatters, the valley filtering effect from intra-valley scattering averages out after integration over the incident angles, a consequence of the time-reversal symmetry that has also dictated the absence of valley filtering in 1D systems. Nevertheless, inter-valley scattering functionalizes the line scatters as valley sources, which can efficiently pump valley current even under the integration over incident angles. We demonstrate these general points with explicit results from $n-n^{-}-n$ junctions in TMDs and graphene, which can have tunable functionality from a filter dominating regime to a source dominating regime, through the adjustment of the band-offset between the $n$ and ${n}^{-}$ regions.
For junctions oriented along a chiral direction, we further show that the valley fluxes are pumped into tunable range of angles well separated from the direction of driving charge current, easing the harvest of the valley polarization in a cross-bar device.


\section{Switchable valley functionalities}

\subsection{Proof of vanishing angle-integrated valley flux in the absence of inter-valley scattering by line scatterers}
\begin{figure*}[h]
 \includegraphics[width=14.5cm,height=11.6cm]{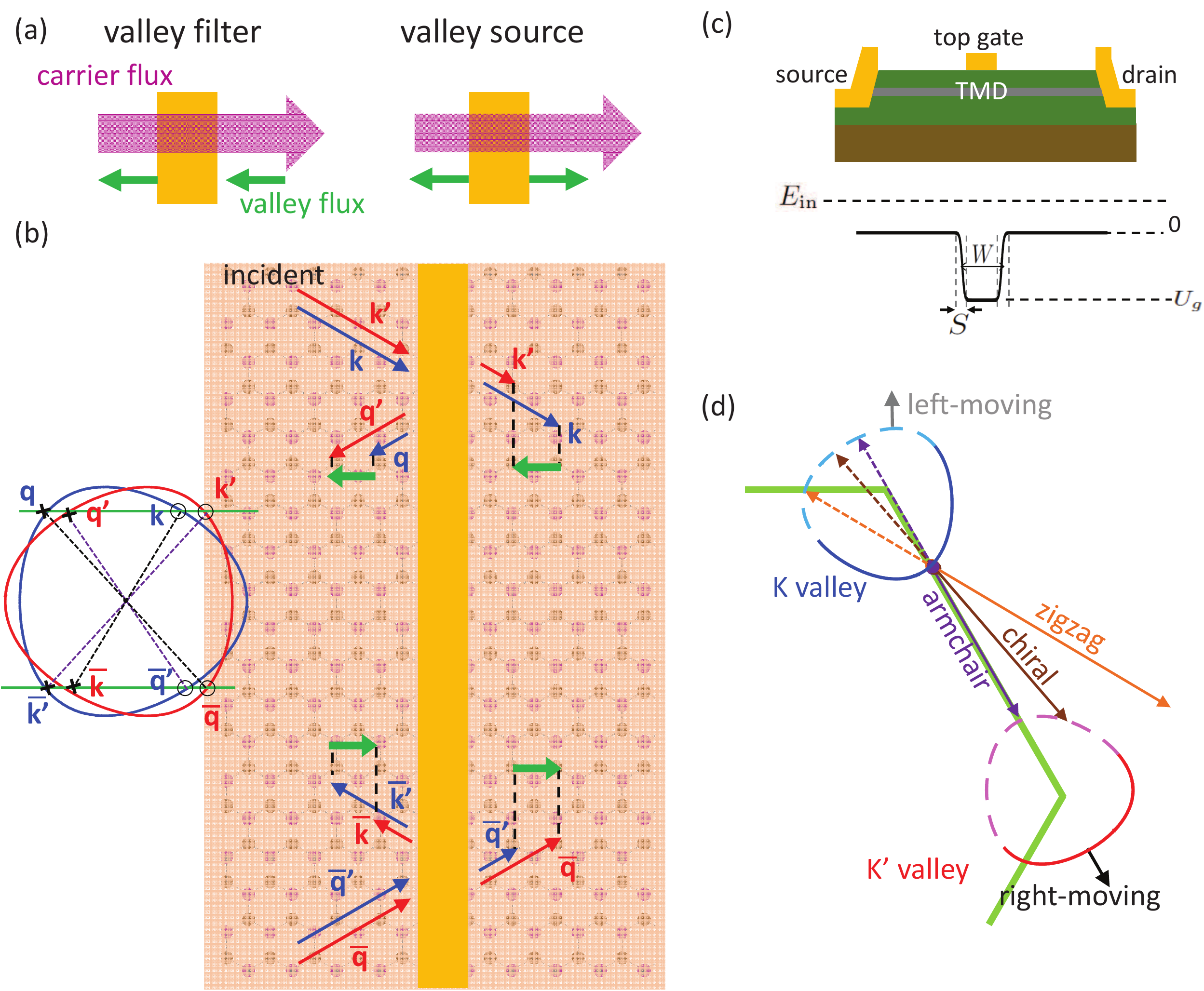} \caption{(color online)
(a):Net valley flows of a valley filter and a valley source. (b): Directional valley filtering for two oblique incidences in 2D producing opposite valley fluxes (see the inset and the texts below). (c): 2D $n-n^{-}-n$ junction formed by deposited gates. (d):The fermi contours around the two valleys in momentum space zoomed in around two corners of the Brillouin zone, displaying how momentum conservation along the interface enables or disenables inter-valley scattering. In (a), the valley filter is more penetrable for in-coming carriers with one valley polarity than the opposite polarity. Without inter-valley scattering, the net valley fluxes flow in the same direction on both sides of the filter (see the green arrows). The valley source however can flip the valley polarity of the in-coming carriers and produces a net out-going valley flux (see oppositely directed green arrows). In (b), we have $\boldsymbol{k}^{\prime},\boldsymbol{q}^{\prime},\bar{\boldsymbol{k}},\bar{\boldsymbol{q}}\in K^{\prime}$ and $\boldsymbol{k}^{},\boldsymbol{q}^{},\bar{\boldsymbol{k}}^{\prime},\bar{\boldsymbol{q}}^{\prime}\in K^{}$, where $\bar{\boldsymbol{\cdot}}$ is the time-reversal of $\boldsymbol{\cdot}$, as indicated in the inset. There the blue and the red closed curves are the fermi contours of the $K$ and $K^{\prime}$ valleys. The incident and the out-going momenta are marked as circles and crosses. Two momenta forming a time-reversal pair are connected by a dashed line. In (c), the junction is described by a potential well. The shape of the potential well is characterized by two length parameters $S_{}$ (smoothing) and $W$ (well width), and a depth parameter $U_{g}$. The dashed horizontal line above indicates the incident energy $E_{\text{in}}$. They are specified in the following calculations.
In (d), the solid- (dashed-)line portions on the fermi contours have right-moving (left-moving) states. An incidence on a right-moving state of the $K$ valley (the spot on the fermi contour of that valley) can be scattered to left-moving state on the same valley (see extended dashed lines from the spot) for all interface orientation. It can be scattered to the $K^{\prime}$ valley only when the interface is not along the zigzag direction, due to momentum conservation along the interface (see the extended solid lines from the spot and their intersecting with the fermi contour of the $K^{\prime}$ valley).}
\label{schematics}
\end{figure*}

We first explicate the general microscopic picture that underlies the vanishing of angle-integrated valley flux in the valley filtering effect of line scatters. We denote the probabilities of scattering the in-coming carrier, injected from the left side of the interface at angle $\theta$ and valley $\tau$, to valley $\tau^{\prime}$, by $R_{\tau\rightarrow\tau^{\prime}}(\theta)$ (as reflection) and $T_{\tau\rightarrow\tau^{\prime}}(\theta)$ (as transmission). The angle-integrated fluxes are given by $\mathcal{R}_{\tau\rightarrow \tau^{\prime}}\equiv(2\pi)^{-1}\int_{}^{}d\theta R_{\tau\rightarrow\tau^{\prime}}(\theta)$ and $\mathcal{T}_{\tau\rightarrow \tau^{\prime}}\equiv(2\pi)^{-1}\int_{}^{}d\theta T_{\tau\rightarrow\tau^{\prime}}(\theta)$ where $(2\pi)^{-1}$ comes as the normalization constant such that $1=\sum_{\tau,\tau^{\prime}\in\{K,K^{\prime}\}}[\mathcal{R}_{\tau\rightarrow \tau^{\prime}}+\mathcal{T}_{\tau\rightarrow \tau^{\prime}}]$. Generally, given a scattering event, one can always find a counterpart event by designating the time-reversal of the in-coming (out-going) momentum of one event to be the out-going (in-coming) momentum of the other. The two events correspond to identical scattering probabilities by respecting time-reversal symmetry.\cite{Newton82book} Explicitly, this implies that given an incident angle $\theta$ and valley $\tau$, one can accordingly find a corresponding angle $\theta^{\prime}$ such that
\begin{align}
\label{ppR}
R_{\tau\rightarrow\tau}(\theta)=R_{\bar{\tau}\rightarrow\bar{\tau}}(\theta^{\prime}),
\end{align} where $\bar{\tau}$ denotes the opposite valley of $\tau$.
For 2D crystals with a time-reversal pair of valleys, we visualize these paired momenta on the fermi contours by the inset of Fig.~\ref{schematics}(b) (see the captions for more details). The consequence of Eq.~(\ref{ppR}) is illustrated in Fig.~\ref{schematics}(b). The upper part of Fig.~\ref{schematics}(b) describes how a valley unpolarized oblique incidence produces a valley current, i.e the valley filtering effect by intra-valley scattering. Its counterpart event is described in the lower part of Fig.~\ref{schematics}(b). The two events produce opposite valley
fluxes, as illustrated in  Fig.~\ref{schematics}(b). Thus, while each intra-valley scattering event has a valley-filtering function, the valley fluxes produced by the pair cancel each other.

On the other hand, directly summing all incident angles leads Eq.~(\ref{ppR}) to
\begin{align}
\label{ref-ang-canl}
\mathcal{R}_{K^{\prime}\rightarrow K^{\prime}}=\mathcal{R}_{K^{}\rightarrow K^{}}.
\end{align} The angle-integrated valley fluxes in the reflection/transmission are defined by
\begin{subequations}
\label{def-angint-vlyflx}
\begin{align}
J^{R/T}_{v}=J^{R/T,v}_{+}+J^{R/T,v}_{-},
\end{align} where
\begin{align}
\label{def-angint-vlyflx-diag}
J^{T,v}_{+}=[\mathcal{T}_{K^{\prime}\rightarrow K^{\prime}}-\mathcal{T}_{K^{}\rightarrow K^{}}],~
J^{R,v}_{+}=-[\mathcal{R}_{K^{\prime}\rightarrow K^{\prime}}-\mathcal{R}_{K^{}\rightarrow K^{}}],
\end{align} and
\begin{align}
\label{def-angint-vlyflx-offdiag}
J^{T,v}_{-}=[\mathcal{T}_{K^{}\rightarrow K^{\prime}}-\mathcal{T}_{K^{\prime}\rightarrow K^{}}],~
J^{R,v}_{-}=-[\mathcal{R}_{K^{}\rightarrow K^{\prime}}-\mathcal{R}_{K^{\prime}\rightarrow K^{}}].
\end{align}
\end{subequations} For valley filters, the contributions $J^{R/T,v}_{-}$ in Eq.~(\ref{def-angint-vlyflx-offdiag}) are zero by definition. Immediately from Eq.~(\ref{ref-ang-canl}) and the conservation law, leading to $\mathcal{T}_{K^{\prime}\rightarrow K^{\prime}}=\mathcal{T}_{K^{}\rightarrow K^{}}$, applied to Eq.~(\ref{def-angint-vlyflx}), the vanishing angle-integrated flux $J^{R/T}_{v}=0$ is proved. Note that Eqs.~(\ref{ppR}) and (\ref{ref-ang-canl}) hold regardless the scatterer acts as a valley filter or valley source. As a result, we ensure that the net valley current under equilibrium condition is zero for both valley filters and sources (see appendix \ref{eqvlyflx}).

\subsection{Angle-resolved valley flux}

\begin{figure*}[h]
 \includegraphics[width=11.5cm,height=8.79412cm]{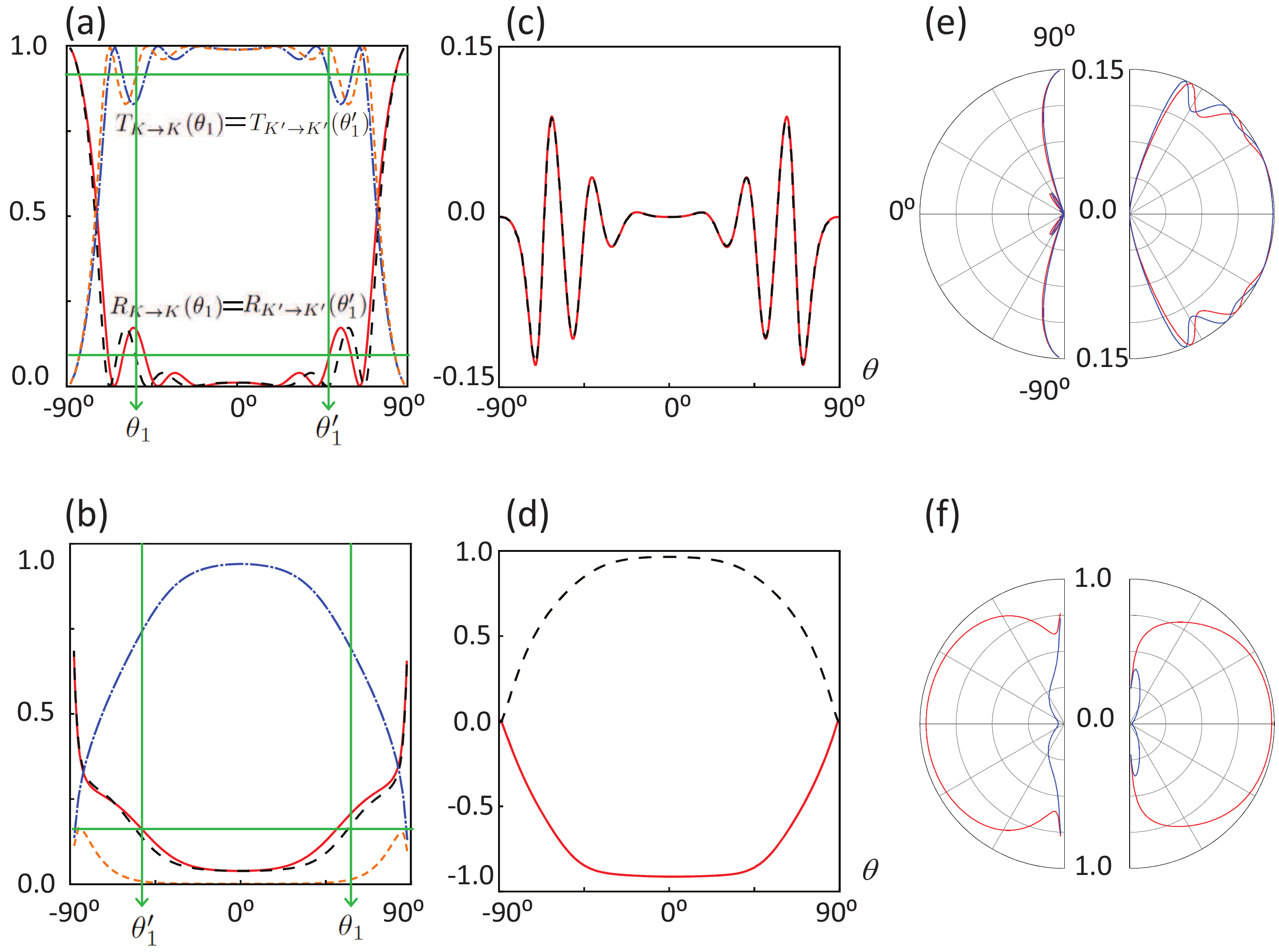}
\caption{(color online) (a) and (b):Reflection coefficients $R_{K^{\prime}\rightarrow K^{\prime}}(\theta)$ (red solid lines), $R_{K^{}\rightarrow K^{}}(\theta)$ (black long-dashed lines) and transmission coefficients $T_{K^{\prime}\rightarrow K^{\prime}}(\theta)$ (blue dash-dot lines), $T_{K^{}\rightarrow K^{}}(\theta)$ (orange short-dashed lines). (c) and (d):The valley fluxes $j^{R}_v(\theta)$ (the red solid line) and $j^{T}_v(\theta)$ (the black long-dashed line), illustrating  valley filters (as (c)) and valley sources (as (d)). (e) and (f): The out-going angle distributions of the reflected (the left semi-circle) and the transmitted (the right semi-circle) fluxes carrying valley $K$ (the blue line) and $K^{\prime}$ (the red line). (a), (c) and (e) are in filter-dominating regime with $U_{g}=-77.4$meV while (b), (d) and (f) are in the source-dominating regime, using $U_{g}=-530$meV. In (a), the intersects between the green vertical and horizontal lines exemplifies the equality shown on the plot's vicinity. These equalities underlie the cancellation of opposite valley fluxes by time-reversal paired scattering events. In (b), only $R_{K\rightarrow K^{}}(\theta_{1})=R_{K^{\prime}\rightarrow K^{\prime}}(\theta^{\prime}_{1})$ is identified. The lack of such equality in transmission coefficients signifies the importance of inter-valley scattering. The incident energy $E_{\text{in}}$ is measured from conduction band bottom, $E_{F}$, by $E_{\text{in}}-E_{F}=44.24$meV. Other well shape parameters are $S_{}=10a$ and $W=110a$, where $a$ is the lattice constant. These parameters are fixed otherwise specified. The slight asymmetry between $\theta^{\prime}_{1}$ and $\theta^{}_{1}$  with respect to $0^{\circ}$ is due to the trigonal warping of the fermi contours.
}
\label{E3S1F3V4_7_30_chgflx_vlyflx_vlypol}
\end{figure*}

Below we illustrate these general relations discussed above by $n-n^{-}-n$ junctions on TMD (schematically shown as Fig.~\ref{schematics}(c)) with the armchair-oriented interface since inter-valley scattering is possible for all incident angles (see Fig.~\ref{schematics}(d) and its captions).
We present in Fig.~\ref{E3S1F3V4_7_30_chgflx_vlyflx_vlypol}(a) and (b) the intra-valley scattering probabilities respectively for small and large band-offsets between the $n$ and $n^{-}$ region (see appendix \ref{TB-method} for the method). Such a junction can be described as a potential well structure. The results at shallow (Fig.~\ref{E3S1F3V4_7_30_chgflx_vlyflx_vlypol}(a)) and deep (Fig.~\ref{E3S1F3V4_7_30_chgflx_vlyflx_vlypol}(b)) well depths both show that for a given incident angle $\theta=\theta_{1}$, one can always find another angle $\theta=\theta^{\prime}_{1}$ (related to $\theta_{1}$ through time-reversal analysis) such that $R_{K\rightarrow K^{}}(\theta_{1})=R_{K^{\prime}\rightarrow K^{\prime}}(\theta^{\prime}_{1})$ (comparing the red solid and the black long-dashed lines in Fig.~\ref{E3S1F3V4_7_30_chgflx_vlyflx_vlypol}(a) and (b)). This witnesses Eq.~(\ref{ppR}). For small band-offsets, the inter-valley scattering becomes negligible, namely, $R_{\tau\rightarrow\bar{\tau}}(\theta_{})\approx0$ and $ T_{\tau\rightarrow\bar{\tau}}(\theta_{})\approx0$. Then by charge conservation ($1=\sum_{\tau'\in\{K,K^{\prime}\}}[R_{\tau\rightarrow \tau^{\prime}}(\theta^{}_{})+T_{\tau\rightarrow \tau^{\prime}}(\theta^{}_{})]$ for all $\tau\in\{K,K^{\prime}\}$), $T_{K\rightarrow K^{}}(\theta_{1})=T_{K^{\prime}\rightarrow K^{\prime}}(\theta^{\prime}_{1})$ also applies (see the blue dash-dot and orange short-dashed lines in Fig.~\ref{E3S1F3V4_7_30_chgflx_vlyflx_vlypol}(a)). These equalities exemplify the situation illustrated in Fig.~\ref{schematics}(b). The valley fluxes on the two sides of the junction are given by
\begin{align}
&j^{T}_v(\theta)=\sum_{\tau\in\{K,K^{\prime}\}}[T_{\tau\rightarrow K^{\prime}}(\theta)-T_{\tau\rightarrow K^{}}(\theta)],\nonumber\\
&j^{R}_v(\theta)=-\sum_{\tau\in\{K,K^{\prime}\}}[R_{\tau\rightarrow K^{\prime}}(\theta)-R_{\tau\rightarrow K^{}}(\theta)],
\end{align}
where the minus sign in $j^{R}_v(\theta)$ accounts for reflected flux flowing oppositely to the transmitted ones. Consequently with a shallow well, they satisfy $j^{R}_v(\theta)=j^{T}_v(\theta)$. The valley flux on the left side thus flows into the junction interface while that on the right side flows outward from the interface, as shown by Fig.~\ref{E3S1F3V4_7_30_chgflx_vlyflx_vlypol}(c). This result shows that the $n-n^{-}-n$ junction with small band-offset behaves as a valley filter (the left panel of Fig.~\ref{schematics}(a)). Note that the impossibility of producing valley current by 1D valley filter is reproduced by zero valley currents at normal incidence, $j^{R}_v(0)=j^{T}_v(0)=0$ (see Fig.~\ref{E3S1F3V4_7_30_chgflx_vlyflx_vlypol}(c)).

At deeper well depth, the inter-valley scattering events then become not ignorable. Although we still have $R_{K\rightarrow K^{}}(\theta_{1})=R_{K^{\prime}\rightarrow K^{\prime}}(\theta^{\prime}_{1})$, charge conservation law together with $R_{K\rightarrow K^{\prime}}(\theta_{1})\ne0$, $R_{K^{\prime}\rightarrow K^{}}(\theta^{\prime}_{1})\ne0$ then leads to $T_{K\rightarrow K^{}}(\theta_{1})\ne T_{K^{\prime}\rightarrow K^{\prime}}(\theta^{\prime}_{1})$ (see the separated range of values between the blue dash-dot and orange short-dashed lines in Fig.~\ref{E3S1F3V4_7_30_chgflx_vlyflx_vlypol}(b)). Under such circumstance, Fig.~\ref{E3S1F3V4_7_30_chgflx_vlyflx_vlypol}(d) shows that the valley fluxes on both sides of the junction flow outward from the interface. This signifies that such $n-n^{-}-n$ junction with sufficient large band-offset works as a valley source (the left panel of Fig.~\ref{schematics}(a)). The out-going fluxes for carrying both valleys are further shown in Fig.~\ref{E3S1F3V4_7_30_chgflx_vlyflx_vlypol}(e) and (f) for small and large band-offsets respectively. Fig.~\ref{E3S1F3V4_7_30_chgflx_vlyflx_vlypol}(e) reveals that the valley filtering effect indeed is more dominant for oblique angles other than those angles close to that of normal incidences, consistent with our argument above. On the contrary, inter-valley scattering induced valley source effect is not constrained to oblique angles, as evidenced by Fig.~\ref{E3S1F3V4_7_30_chgflx_vlyflx_vlypol}(f).

\subsection{Angle-integrated valley flux}

\subsubsection{Interface along armchair direction}

\begin{figure*}[h]
 \includegraphics[width=14.5cm,height=10.79412cm]{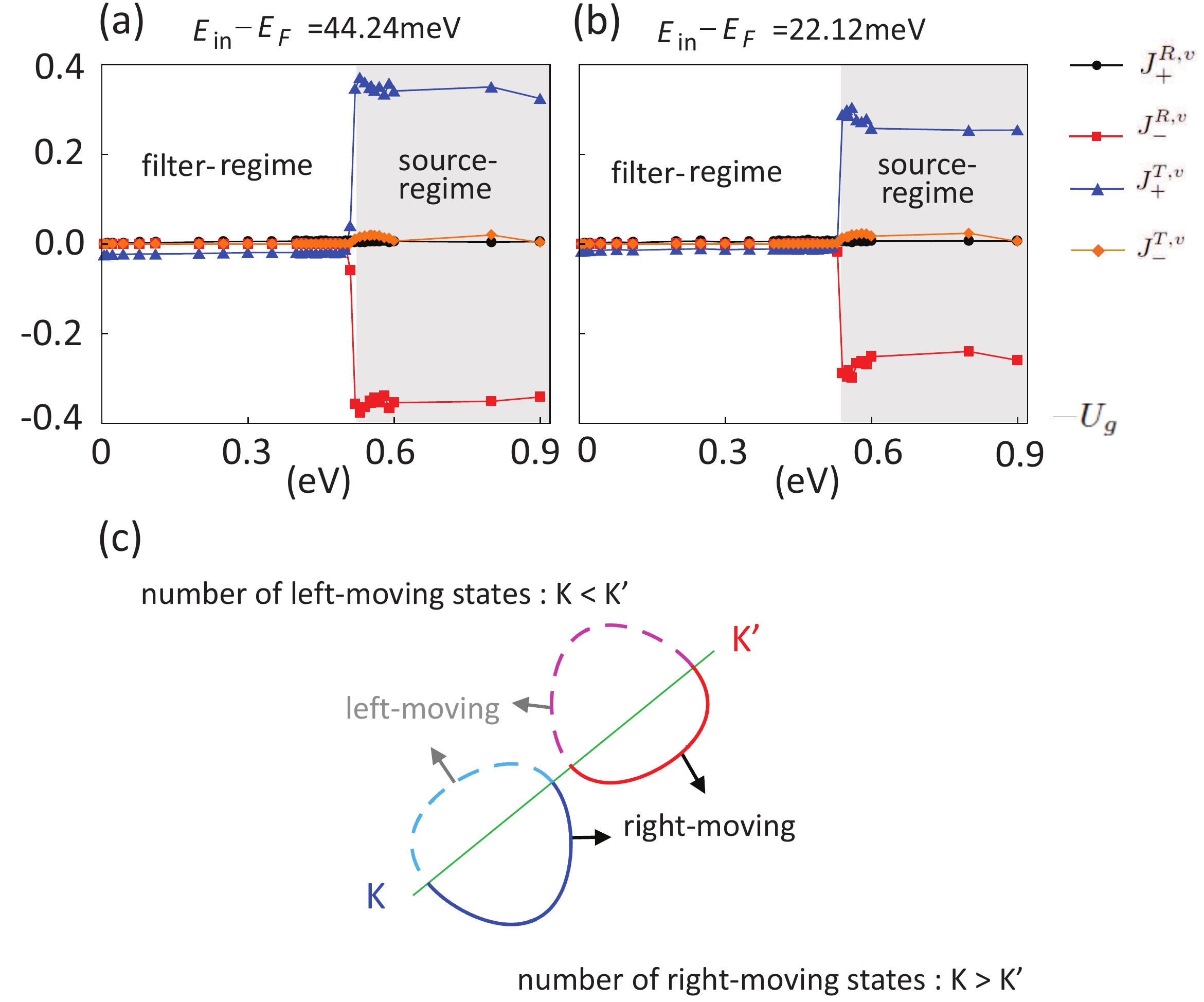}
\caption{(color online)(a) and (b):The angle-integrated contributions to the valley fluxes, $J^{R,v}_{+}$ (the black circles), $J^{R,v}_{-}$ (the red squares), $J^{T,v}_{+}$ (the blue triangles), $J^{T,v}_{-}$ (the orange diamonds). (c):Illustration of the trigonal warping effect on valley injection without scattering. (b) only differs from (a) by the incident energy, as indicated above these plots. Both (a) and (b) reveal the cancellation of time-reversal related backscattering can lead to vanishing valley flux, as $J^{R,v}_{+}$ remain zero independently of the well depth. The disappearing of angle-integrated valley flux in the filter-dominating regime and the rise of it in the source-dominating-regime (the gray shaded areas) are clearly displayed. In (c), the solid (dashed) portion on $K^{}$ is obviously longer (shorter) than that in $K^{\prime}$, giving rise to valley polarization of the injected carriers due the trigonal warping effect without scattering.}
\label{E2_3_4S1_2_3F2_3_5V7_vlypol}
\end{figure*}

\begin{figure*}[h]
 \includegraphics[width=14.5556cm,height=4.38662cm]{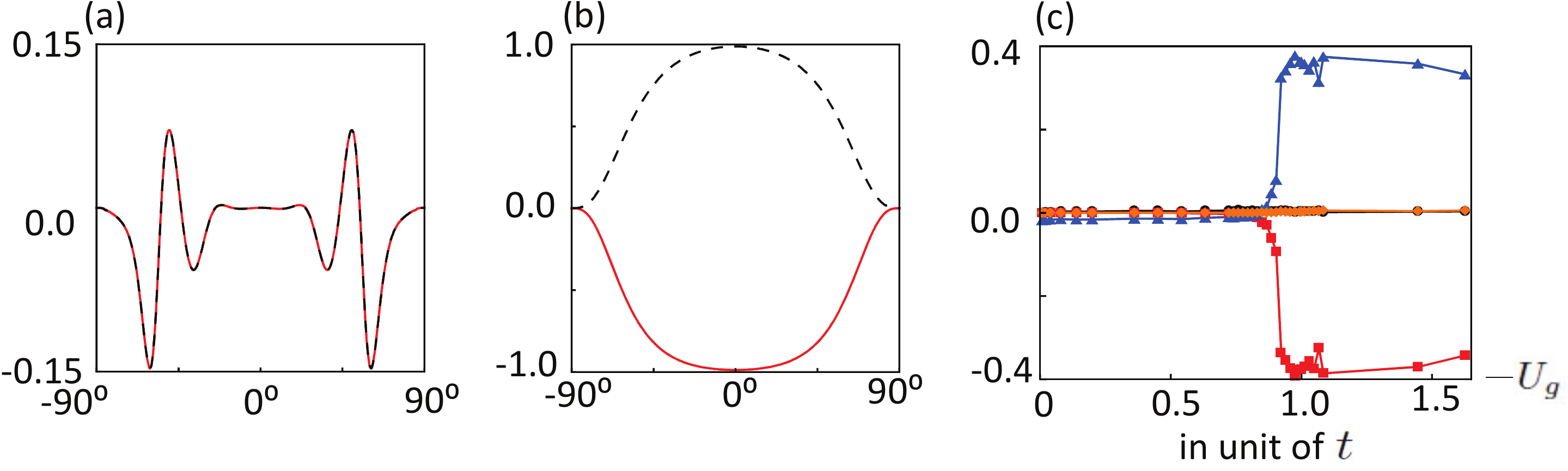}
\caption{Results for graphene with the interface oriented in the armchair direction, verifying the general points raised in the main text. In (a) and (b) we show the valley fluxes $j^{R}_v(\theta)$ (the red solid line) and $j^{T}_v(\theta)$ (the black long-dashed line). (a) is in the filter-dominating regime with $U_{g}=-0.08\left\vert t\right\vert$ and (b), in the source-dominating regime, is with $U_{g}=-0.97\left\vert t\right\vert$ for $t\approx-2.8$eV being the hopping between neighboring sublattices.\cite{Neto09109} (c) shows various contributions to angle-integrated valley fluxes, $J^{R,v}_{+}$ (the black circles), $J^{R,v}_{-}$ (the red squares), $J^{T,v}_{+}$ (the blue triangles), $J^{T,v}_{-}$ (the orange diamonds). In (a),(b) and (c), we use $E_{\text{in}}-E_{F}=0.08\left\vert t\right\vert$. Other parameters are $S_{b}=10a$ and $W=260a$ for $a$ the lattice constant of graphene. }
\label{graphene}
\end{figure*}

The different contributions to the angle-integrated valley fluxes are further presented as different symbols in Fig.~\ref{E2_3_4S1_2_3F2_3_5V7_vlypol}(a) and (b) respectively for higher (sufficient for visible trigonal warping) and lower incident energies, as a function of the well depth. The cancellation of the reflected valley fluxes discussed previously, Eq.~(\ref{ref-ang-canl}), is a general phenomena and is expected to be independent of the details of the scatterers. This is witnessed by $J^{R,v}_{+}$ being a vanishing constant (see Eq.~(\ref{def-angint-vlyflx-diag})) in both Fig.~\ref{E2_3_4S1_2_3F2_3_5V7_vlypol}(a) and (b) (the black circles), independent of the well depths. A small valley flux contributed by $J^{T,v}_{+}$ readily appears at zero well depth (see the values of the blue triangles for $U_{g}=0$ in Fig.~\ref{E2_3_4S1_2_3F2_3_5V7_vlypol}(a) and (b)). It is due to more injection in $K^{\prime}$ than in $K$ (see Fig.~\ref{E2_3_4S1_2_3F2_3_5V7_vlypol}(c) and its captions) without scattering and such difference is reduced by shifting the incident energy closer to the band edge (comparing the values of the blue triangles for $U_{g}=0$ in Fig.~\ref{E2_3_4S1_2_3F2_3_5V7_vlypol}(a) and (b)). Apart from this, in the filter-dominating regime (shallower well depth with negligible $J^{R/T,v}_{-}$), the angle-integrated valley fluxes excluding $J^{T,v}_{+}$ is very small. This verifies the above general analysis for valley filters. Tuning of the well depth from filter-dominating to the source-dominating regime is accompanied by the rise of the importance of inter-valley scattering, $J^{R,v}_{-}$ (the red squares) and $J^{T,v}_{-}$ (the orange diamonds) around $U_{g}\approx-550$meV. The consequent rise of $J^{T,v}_{+}$ (the blue triangles) is due to the charge conservation with sizable contribution from inter-valley reflection $J^{R,v}_{-}$. These results demonstrate that the rise of angle-integrated valley flux are from the actions of a valley source, other than a valley filter. The same kinds of investigations into the valleytronics of 2D $n-n^{-}-n$ junctions on TMD are also carried out for graphene (see Fig.~\ref{graphene}). The results there show that the angle-integrated valley flux is zero in the filter regime and the band-offset can be tuned to switch the valley functionalities between valley filters and valley sources, reaching the same conclusion for 2D $n-n^{-}-n$ junctions on TMD.

\subsubsection{Interface along chiral direction}

\begin{figure*}[h]
 \includegraphics[width=14.5cm,height=8.5cm]{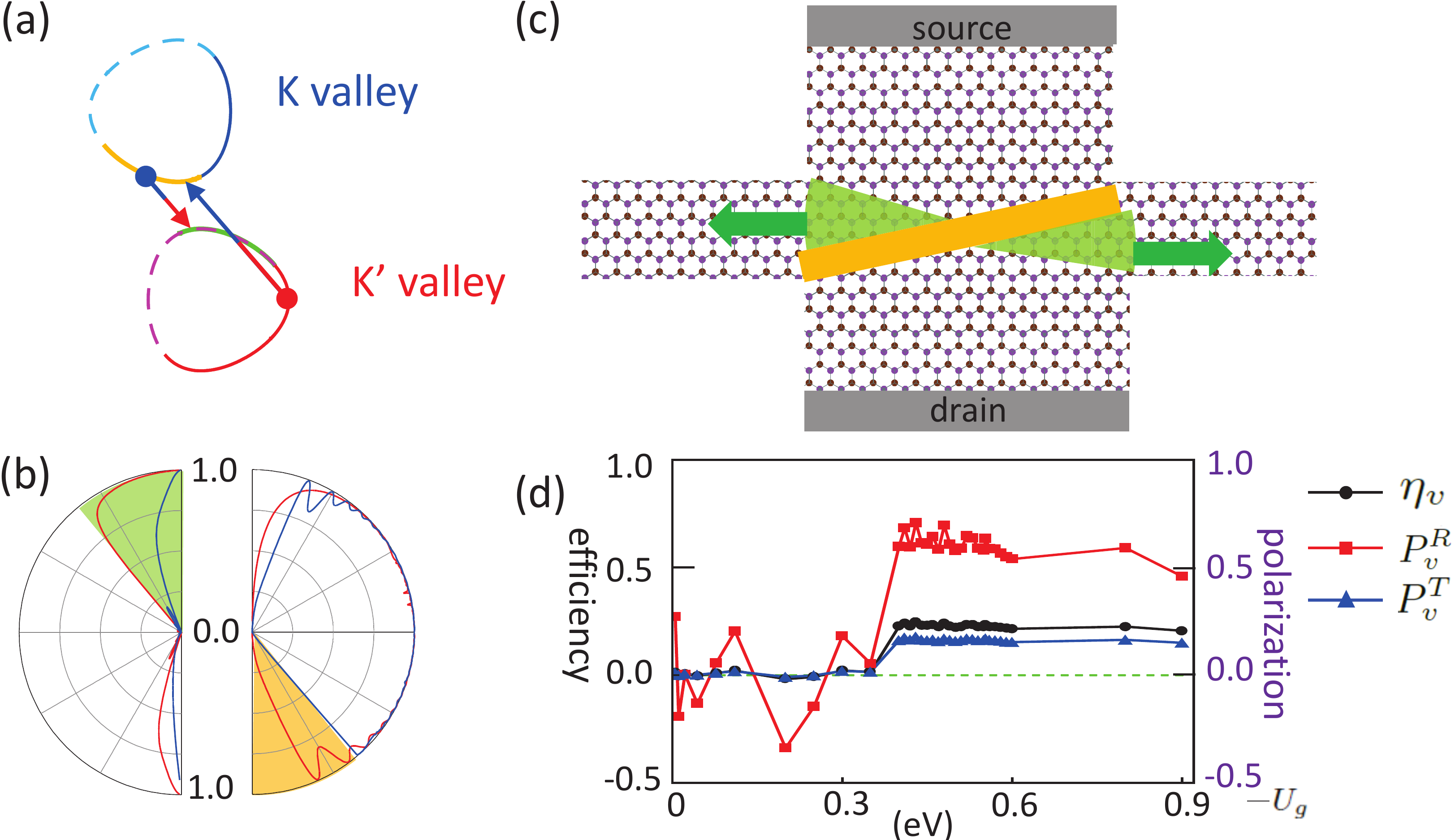}
\caption{(color online) Chiral oriented interface results in separation between the directions of the valley and charge fluxes. (a):Portions (differently colored) on the fermi contours upon which the incident carrier is allowed to be scattered into the opposite valley. (b):Out-going fluxes carrying valleys $K$ (blue) and $K^{\prime}$ (red), showing the deviation of the direction of valley flux from the biased direction. (c):A possible design of valley antenna by cross-bar geometry, where the bias is applied in the vertical direction while the valley fluxes are collected by horizontal electrodes. (d):The efficiency, $\eta_{v}$ (the black disks), the polarizations of the reflected ($P^{R}_{v}$ the red squares) and the transmitted ($P^{T}_{v}$ the blue triangles) fluxes within the collectable angular ranges. In (a), incidence (as the right moving states) in $K$ ($K^{\prime}$) is allowed to be scattered to the green (light-orange) sector in $K^{\prime}$ (K). The green and the light-orange colored portions in (a) are correspondingly shaded in (b). In (b) and (d), the incident energy is $E_{\text{in}}-E_{F}=110.6$meV while $U_{g}=-400$meV is used in (b). Here we take the (1,2) direction. }
\label{OVchglx_vlypol_E1_3_5_S1_2_3_F2_3_5}
\end{figure*}

The above discussion has shown that the effectiveness of the valley source relies on the inter-valley scattering. To better harness these valley fluxes, one then wishes to separate the valley flow from the charge flow. In Fig.~\ref{E3S1F3V4_7_30_chgflx_vlyflx_vlypol}(f) (as a valley source), the valley fluxes are distributed over a range of angles of $180^{\circ}$, predetermined by the interface orientation being the armchair one. By orientating the 2D junction along a chiral direction, the angular range of inter-valley scattering becomes concentrated and deviated from the bias direction (see Fig.~\ref{OVchglx_vlypol_E1_3_5_S1_2_3_F2_3_5}(a)). This provides the possibility to separate the valley flux from the charge flux induced by the bias. The results are exemplified in Fig.~\ref{OVchglx_vlypol_E1_3_5_S1_2_3_F2_3_5}(b), showing that the valley fluxes are focused within a certain range of angles, oriented away from the bias direction (see the yellow/green shaded areas in Fig.~\ref{OVchglx_vlypol_E1_3_5_S1_2_3_F2_3_5}(b)). The subsequent valley fluxes in oblique directions can be collected by additional electrodes that extend perpendicularly to the biased direction, as the cross-bar sketched in Fig.~\ref{OVchglx_vlypol_E1_3_5_S1_2_3_F2_3_5}(c).

\begin{figure*}[h]
 \includegraphics[width=13.7851cm,height=4.38662cm]{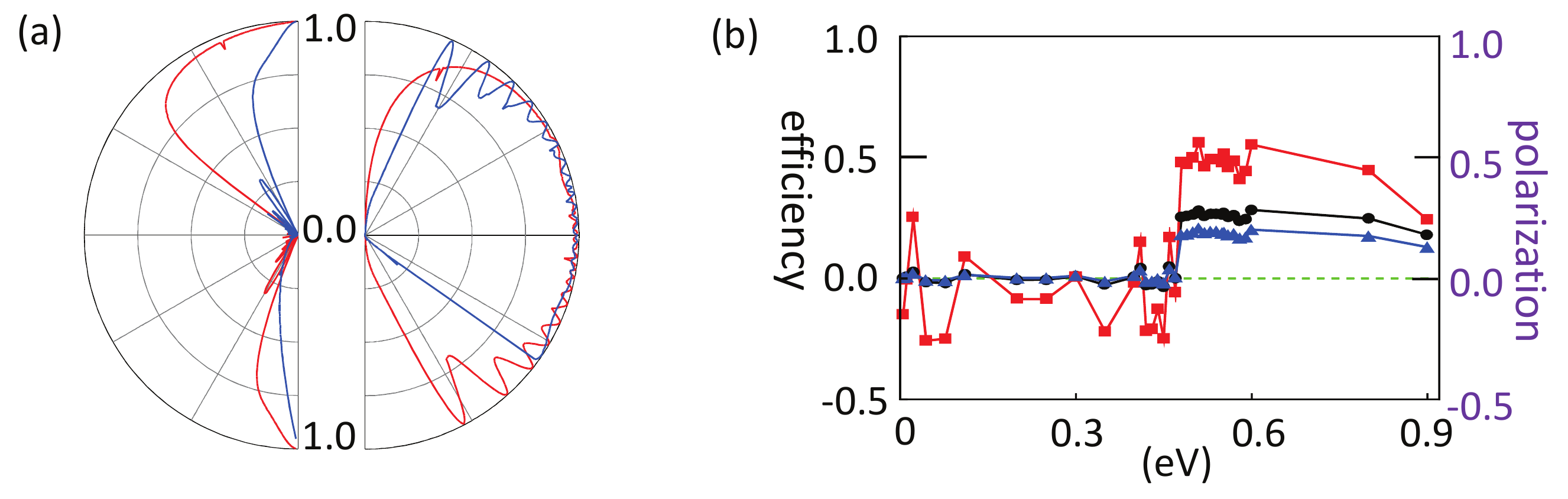}
\caption{Results from a cross-bar valley source and collector using another chiral direction, (2,3), based on MoS${}_{2}$. In (a), we show for this chiral direction the out-going fluxes carrying valleys $K$ (blue) and $K^{\prime}$ (red), similar to the example for the chiral direction used in the main text. In (b), the performance of the cross-bar device is measured with the same quantities, namely, the efficiency, $\eta_{v}$ (the black disks), the polarizations of the reflected ($P^{R}_{v}$ the red squares) and the transmitted ($P^{T}_{v}$ the blue triangles) fluxes within the collectable angular ranges with $\alpha=\pi/6$. The incident energy is $E_{\text{in}}-E_{F}=77.42$meV.}
\label{MoS2_2_3}
\end{figure*}

Below we assess the performance of the above scenario of generating and collecting valley fluxes.
The longitudinal direction of the collecting electrodes makes an angle $\alpha$ with the interface orientation of the junction. The collected fluxes carrying valley $\tau^{\prime}$ obtained from incidence at valley $\tau$ are thus given by $\hat{\mathcal{R}}_{\tau\rightarrow\tau^{\prime}}=\mathcal{N}\int_{-\alpha}^{\pi/2}d\theta R_{\tau\rightarrow\tau^{\prime}}(\theta)$ and $\hat{\mathcal{T}}_{\tau\rightarrow\tau^{\prime}}=\mathcal{N}\int_{-\pi/2}^{\alpha}d\theta T_{\tau\rightarrow\tau^{\prime}}(\theta)$ for the reflected and the transmitted beams respectively. The collected valley (charge) fluxes are then defined by
$\hat{J}^{T}_{v/c}=\sum_{\tau\in\{K,K^{\prime}\}}[\hat{\mathcal{T}}_{\tau\rightarrow K^{\prime}}\mp\hat{\mathcal{T}}_{\tau\rightarrow K^{}}]$ and
$\hat{J}^{R}_{v/c}=-\sum_{\tau\in\{K,K^{\prime}\}}[\hat{\mathcal{R}}_{\tau\rightarrow K^{\prime}}\mp\hat{\mathcal{R}}_{\tau\rightarrow K^{}}]$, where the subscript "$v/c$" stands for valley/charge and the lower sign is for the charge flux. The efficiency of generating collectable valley fluxes relative to passing charge current is defined by,
\begin{align}
\eta_{v}=\frac{\hat{J}^{T}_{v}-\hat{J}^{R}_{v}}
{\sum_{\tau,\tau^{\prime}\in\{K,K^{\prime}\}}\mathcal{T}_{\tau\rightarrow\tau^{\prime}}},
\end{align} where the numerator stands for the collected total out-going valley flux while the denominator is the net charge transmission. This is plotted as the black disks in Fig.~\ref{OVchglx_vlypol_E1_3_5_S1_2_3_F2_3_5}(d), whose values are calibrated on the left black vertical axis. The quality of the collected valley fluxes can be quantified by the valley polarizations,
\begin{align}
P^{T/R}_{v}=\frac{\hat{J}^{T/R}_{v}}{\hat{J}^{T/R}_{c}}.
\end{align} The valley polarizations of the collected fluxes as $P^{R}_{v}$ and $P^{T}_{v}$ are plotted in Fig.~\ref{OVchglx_vlypol_E1_3_5_S1_2_3_F2_3_5}(d), calibrated by the purple vertical axis on the right. The performance evaluation is done with $\alpha=\pi/6$. When the well is tuned into the source regime (here around $-U_{g}=400$meV), one of the collecting electrode can receive fluxes with valley polarization over 60\%. Such considerable content of valley polarization without interference from bias driven charge current may be useful for further applications.  The effect of concentrating valley currents into angular intervals separated from the bias direction is not restricted to the specific chiral direction used in Fig.~\ref{OVchglx_vlypol_E1_3_5_S1_2_3_F2_3_5}. We provide another example using a different chiral orientation showing similar results in Fig.~\ref{MoS2_2_3}, evidencing its generality. Note that we do not further discuss the case of zigzag-oriented interface since no inter-valley scattering can be induced by such interface (see Fig.~\ref{schematics}(d)).


\section{Conclusions}

Here we summarize our main findings and their general implications on valleytronics for 2D materials. (i):2D valley filters, operated by valley-dependent intra-valley scattering, give vanishing angle-integrated valley flux, due to the nature of the time-reversal pairing between the two valleys. On the other hand, 2D valley sources, operated by significant inter-valley scattering, can generate sizable angle-integrated valley flux. (ii):The gate-tunable band-offset of 2D $n-n^{-}-n$ junctions adjusts the relative importance between intra-valley and inter-valley scattering. The functionalities of valley filters and valley sources can thus be switched by changing the band-offset of 2D $n-n^{-}-n$ junctions. These valley-related effects are demonstrated in both TMDs with the finite bandgap and in gapless graphene. (iii):The orientation of 2D $n-n^{-}-n$ junctions determine the incident angles by which inter-valley scattering is allowed. Therefore, by orienting the junction in chiral directions, the directions of valley fluxes can be deviated from the biased direction of charge fluxes. Combining such 2D junctions with extra electrodes to form cross-bar geometries, the valley fluxes generated by the valley sources can be separately harvested by the collecting electrodes. The valley filter and valley source effects of these 2D $n-n^{-}-n$ junctions are also anticipated in other materials with time-reversal paired valleys. The investigations carried out here also show that the benifits of using 2D materials as platforms for valleytronic operations are not only provided by the easily accessible valleys but also by the nature of valleys as momentum index whose 2D nature can be manipulated.

\begin{acknowledgements}
This work is supported by the Croucher Foundation under the Croucher Innovation Award, the RGC (HKU9/CRF/13G) and UGC (AoE/P-04/08) of HKSAR, and HKU ORA.
This research is conducted in part using the HKU ITS research computing facilities that are supported in part by the Hong Kong UGC Special Equipment Grant (SEG HKU09).
\end{acknowledgements}

%
%
%
%
%

\appendix


\section{Equilibrium valley current}
\label{eqvlyflx}
Here we extend our analysis in the main text for deducing that the vanishing of the angle-integrated valley flux for 2D valley filters. We apply similar analysis to inspect the net valley flux under the equilibrium condition, namely, the condition where no charge current flows, in order to see if it gives sensible conclusion. Below we show that the net valley flux under equilibrium is reasonably zero.

The equilibrium situation is attained by injecting equal amount of carriers from both sides of the interface such that the net charge flux is zero. We denote the probabilities of scattering the in-coming carrier, injected from the left/right side of the interface at angle $\theta$ and valley $\tau$, to valley $\tau^{\prime}$, by $R^{1/2}_{\tau\rightarrow\tau^{\prime}}(\theta)$ (as reflection) and $T^{1/2}_{\tau\rightarrow\tau^{\prime}}(\theta)$ (as transmission). Let us consider a specific injection of a carrier from the left side at valley $\tau$ with angle $\theta^{1}_{a,\tau}$. This in-coming momentum is denoted by $\boldsymbol{k}^{+,a}_{\tau}$.
According to momentum conservation, the out-going momenta can be explicitly determined. We denote the out-going momenta reached by forward/backward intra-valley and inter-valley scattering by $\boldsymbol{k}^{+/-,a}_{\tau}$ and $\boldsymbol{k}^{+/-,a}_{\bar{\tau}}$ respectively. Note that due to the obliqueness of the incidence, the momenta $\boldsymbol{k}^{+,a}_{\tau}$ and $\boldsymbol{k}^{-,a}_{\bar{\tau}}$  no longer make a time-reversal pair. By time reversal symmetry, the momentum $\boldsymbol{k}^{\pm,a}_{\tau^{\prime}}$ is paired with $\boldsymbol{k}^{\mp,b}_{\bar{\tau}^{\prime}}$ for $\tau^{\prime}\in\{K,K^{\prime}\}$. Automatically, the momenta denoted by $\boldsymbol{k}^{\mp,b}_{\tau^{\prime}}$'s share the same projection along the interface. Denoting the associated angle of a momentum $\boldsymbol{k}^{\pm,a/b}_{\tau^{\prime}}$ by $\theta^{\pm,a/b}_{\tau}$, the time-reversal symmetry applied to the intra-valley backscattering reads
\begin{subequations}
\label{TR-ref}
\begin{align}
R^{1/2}_{\tau\rightarrow\tau}(\theta^{+/-,a}_{\tau})=R^{1/2}_{\bar{\tau}\rightarrow\bar{\tau}}(\theta^{+/-,b}_{\bar{\tau}}),
\end{align} and
\begin{align}
T^{1}_{\tau\rightarrow\tau^{\prime}}(\theta^{+,a/b}_{\tau})=T^{2}_{\bar{\tau}^{\prime}\rightarrow\bar{\tau}}(\theta^{-,b/a}_{\bar{\tau}^{\prime}}).
\end{align}
\end{subequations}
for forward scattering. These scattering probabilities are subjected to normalization, namely,
$1=\sum_{\tau^{\prime}\in\{K,K^{\prime}\}}[R^{1/2}_{\tau\rightarrow\tau^{\prime}}(\theta^{+/-,a}_{\tau})+T^{1/2}_{\tau\rightarrow\tau^{\prime}}(\theta^{+/-,a}_{\tau})]
=\sum_{\tau^{\prime}\in\{K,K^{\prime}\}}[R^{1/2}_{\tau\rightarrow\tau^{\prime}}(\theta^{+/-,b}_{\tau})+T^{1/2}_{\tau\rightarrow\tau^{\prime}}(\theta^{+/-,b}_{\tau})]$ for $\tau\in\{K,K^{\prime}\}$.

By simultaneously injecting carriers with momenta $\boldsymbol{k}^{+,a}_{K}$, $\boldsymbol{k}^{+,a}_{K^{\prime}}$, $\boldsymbol{k}^{+,b}_{K}$, $\boldsymbol{k}^{+,b}_{K^{\prime}}$ from the left and momenta $\boldsymbol{k}^{-,a}_{K}$, $\boldsymbol{k}^{-,a}_{K^{\prime}}$, $\boldsymbol{k}^{-,b}_{K}$, $\boldsymbol{k}^{-,b}_{K^{\prime}}$ from the right, the subsequent flux that carriers valley $K^{\prime}$ flowing on the left side reads
\begin{align}
\label{valley-spccrnt}
J^{1}_{K^{\prime}}=&\sum_{\sigma\in\{a,b\}}\left[\left(1-R^{1}_{K^{\prime}\rightarrow K^{\prime}}(\theta^{+,\sigma}_{K^{\prime}})\right)
-R^{1}_{K^{}\rightarrow K^{\prime}}(\theta^{+,\sigma}_{K})\right]
\nonumber\\&
-\sum_{\sigma\in\{a,b\}}\sum_{\tau\in\{K,K^{\prime}\}}T^{2}_{\tau\rightarrow K^{\prime}}(\theta^{-,\sigma}_{\tau}).
\end{align} The first term, $\left(1-R^{1}_{K^{\prime}\rightarrow K^{\prime}}(\theta^{+,\sigma}_{K^{\prime}})\right)$, in Eq.~(\ref{valley-spccrnt}) describes that the injection
at valley $K^{\prime}$ from the left side is backscattered to the same valley, resulting in a net valley current that is the injected one subtracting
the reflected one. The second term $-R^{1}_{K^{}\rightarrow K^{\prime}}(\theta^{+,\sigma}_{K})$ contributing to the flux carrying valley $K^{\prime}$ comes from the inter-valley backscattering of an injection from the left side at valley $K$. The last term $T^{2}_{\tau\rightarrow K^{\prime}}(\theta^{-,\sigma}_{\tau})$ is contributed by the scattering of the injected the carriers at valley $\tau$ from the right side to valley $K^{\prime}$ on the left side. The minus sign stands for flowing toward the left. Similar expressions can be obtained for the net flux that carriers a particular valley lowing on either side of the interface.

Applying Eq.~(\ref{TR-ref}) to Eq.~(\ref{valley-spccrnt}) leads to $J^{1}_{K^{\prime}}=0$ and similarly all other valley-carrying fluxes on either side
of the interface vanish. One can extend the above procedure to include all possible incident angles and the angle-integrated valley fluxes at equilibrium condition
are just zero. The time-reversal symmetry, Eq.~(\ref{TR-ref}), manifested for discrete set of incident angles is also confirmed numerically. Note that in obtaining Eq.~(\ref{TR-ref}) and the consequences $J^{i}_{\tau}=0$ for $i\in\{1,2\}$ and $\tau\in\{K,K^{\prime}\}$, both the intra- and inter-valley scattering have been taken into account.

\section{Methodology}
\label{TB-method}

\subsection{General approach}
\label{appxsec_scs}

A general two-dimensional system is described by the following tight-binding model,
\begin{align}
H_{}=  \sum_{\boldsymbol{r}}\sum_{\boldsymbol{\delta}\in{D}}\sum_{\alpha,\beta}
c_{\boldsymbol{r},\alpha}^{\dagger}h_{\alpha,\beta}(\boldsymbol{\delta})c_{\boldsymbol{r}+\boldsymbol{\delta},\beta}^{} +\sum_{\boldsymbol{r}}\sum_{\alpha}c_{\boldsymbol{r},\alpha}^{\dagger}U_{\alpha}(\boldsymbol{r})c_{\boldsymbol{r},\alpha}^{},
\label{TB-H0}
\end{align}
 where $\boldsymbol{r}$ denotes the position vectors of the lattice
sites and $\alpha,\beta$ the internal orbitals. The displacement
vectors are summed over neighbouring lattice sites, $D=\{0,\pm\boldsymbol{a}_{1},\pm\boldsymbol{a}_{2},\pm(\boldsymbol{a}_{1}+\boldsymbol{a}_{2}),\pm(\boldsymbol{a}_{1}-\boldsymbol{a}_{2})\}$,
where $\boldsymbol{a}_{1}$ and $\boldsymbol{a}_{2}$ are the two
lattice vectors. The operator $c_{\boldsymbol{r},\alpha}^{\dagger}$
($c_{\boldsymbol{r},\alpha}^{}$) then creates (annihilates) an electron
on the oribtal $\alpha$ belonging to the lattice site at $\boldsymbol{r}$.
The energy matrix, $h_{\alpha,\beta}(\boldsymbol{\delta})$, describes
the on-site orbital energies ($\alpha=\beta$) and their mixing ($\alpha\ne\beta$)
for $\boldsymbol{\delta}=0$ and hopping between neighbouring lattice sites for
$\boldsymbol{\delta}\in{D}$ with $\boldsymbol{\delta}\ne0$. The hermiticity of the Hamiltonian
is given by $h_{\alpha,\beta}(\boldsymbol{\delta})=\left[h_{\beta,\alpha}(-\boldsymbol{\delta})\right]^{*}$.
An external potential is applied, specified by $U_{\alpha}(\boldsymbol{r})$,
to shift the energy of orbital $\alpha$ residing on the lattice site
at $\boldsymbol{r}$. The external potential $U_{\alpha}(\boldsymbol{r})$ is zero except for a certain region
called the scattering area. For numerical calculations performed here, the parameters in Eq.~(\ref{TB-H0}) are taken from the ab-inito justified three-band
tight-binding models fitted to a series of transition-metal dichalcogenides
in Ref. {[}\onlinecite{Liu13085433}{]} for $\text{MoS}_{2}$.

A widely applied approach to derive reflection/transmission coefficients
for nanoribbons/tubes (as quasi-one-dimensional systems) intercepted in the middle by a scattering area
is based on a mode-matching technique, initially prescribed for square
lattice\cite{Ando918017} and later generalized to arbitrary lattice.\cite{Khomyakov05035450}
For generic two-dimensional systems, it is widely assumed that the potential along a chiral direction
shows translational invariance (see Refs. [4,8-12] in the main text). Such invariance
can be utilized to transform the two-dimensional scattering problem
into a quasi-one-dimensional problem. Therefore, in principle, the
methods developed in Refs.~[\onlinecite{Ando918017,Khomyakov05035450}],
can be applied after the transformation. For pedagogical purposes, we provide an alternative route of deriving the
required quantities following the line of thoughts used in quantum
mechanics textbooks. We point out its benefits for dealing with the scattering problem in which
the two-dimensional nature of the involved momenta is important.

Below, we first introduce effective lattices useful for transforming the problem at hand to a quasi-one-dimensional equivalence in Sec.~\ref{eff-latt}. The approaches used in Refs.~[\onlinecite{Ando918017,Khomyakov05035450}]
rely on contructing Bloch matrices from a set of eigenvalues and non-orthogonal
eigenvectors. They are obtained through solving the tight-binding
equation by imposing the Bloch symmetry to determine the properties
of the involved modes. In Sec.~\ref{scatt-stat}, we show that the properties of the involved
modes can also be directly obtained from the eigen-equation leading
to the dispersion relation. The physical meanings of these eigenvalues
and eigenvectors can be directly interpreted from this precursor of dispersion relation. The
non-orthogonality of the eigenvectors also naturally appear there. Instead
of getting the scattering states by contructing
the Bloch matrices, we closely follow the usual textbook approach
of matching wavefunctions at the boundaries of the scattering area.
From the scattering states, the next step is to obtain the transmission
and the reflection coefficients. In Refs. {[}\onlinecite{Ando918017,Khomyakov05035450}{]},
the physical transmission is obtained by normalizing the generalized
transmission matrix elements with respect to the current, defined
from the Bloch velocities in terms of the above mentioned eigenvalues and
eigenvectors. Here we wish to remind that the fulfillment of current
conservation testifies the scattering states as the eigenstates. In Sec.~\ref{crnt-ref-trsm}, the reflection and transmission coefficients are then subsequently
deduced from the current conservation law, in which the expression
of the Bloch velocity naturally emerges.



\subsubsection{Effective lattice for interface oriented at a chiral direction}
\label{eff-latt}

We consider the situation in which the scattering area is formed by
virtually cutting a nanoribbon from the two-dimensional crystal. The longitudinal direction of the ``ribbon''
is described by a chiral vector,
\begin{align}
\boldsymbol{A}_{2}=n_{21}\boldsymbol{a}_{1}+n_{22}\boldsymbol{a}_{2},\label{A2def}
\end{align}
 where the integers $n_{21}$ and $n_{22}$ do not have common divisor,
except one. The transverse direction of the ``ribbon''
defines another vector, denoted by
\begin{align}
\boldsymbol{A}_{1}=n_{11}\boldsymbol{a}_{1}+n_{12}\boldsymbol{a}_{2},\label{A1def}
\end{align}
 where $n_{11}$ and $n_{12}$ are integers that do not have common
divisor. The armchair orientation is then done by using $(n_{21},n_{22})=(1,1)$. Other chiral directions
follow different choices of $(n_{21},n_{22})$ specified in the relevant parts of discussions.
We use $(n_{11},n_{12})=(1,-1)$ for the transverse direction in all cases to shrink the number of original lattice sites
in an effective lattice point (see explanation for effective lattice below).

The chiral orientation of the interface can thus be handled
by defining a new effective lattice whose lattice points are composed
of the original lattice sites. The lattice vectors for the effective
lattice are thus given by Eqs.~(\ref{A2def}) and (\ref{A1def}).
The Hamiltonian Eq.~(\ref{TB-H0}) rewritten in the new basis becomes
\begin{align}
\label{TB-H0-eff}
H_{}=  \sum_{\tilde{\boldsymbol{r}}}\sum_{\tilde{\boldsymbol{\delta}}\in{\tilde{D}}}
\sum_{\tilde{\alpha},\tilde{\beta}}
\tilde{c}_{\tilde{\boldsymbol{r}},\tilde{\alpha}}^{\dagger}\tilde{h}_{\tilde{\alpha},\tilde{\beta}}(\tilde{\boldsymbol{\delta}})
\tilde{c}_{\tilde{\boldsymbol{r}}+\boldsymbol{\delta},\tilde{\beta}}^{}
  +\sum_{\tilde{\boldsymbol{r}}}\sum_{\tilde{\alpha}}
 \tilde{c}_{\tilde{\boldsymbol{r}},\tilde{\alpha}}^{\dagger}\tilde{U}_{\tilde{\alpha}}(\tilde{\boldsymbol{r}})
 \tilde{c}_{\tilde{\boldsymbol{r}},\tilde{\alpha}}^{}
\end{align}
 where the vectors $\tilde{\boldsymbol{r}}$ now enumerate the positions
of the sites of the effective lattice and $\tilde{\alpha},\tilde{\beta}$
the effective orbital, comprising the original lattice position and
its orbital. The set of vectors connecting neighbouring sites on the
effective lattice is given by $\tilde{D}=\{0,\pm\boldsymbol{A}_{1},\pm\boldsymbol{A}_{2},\pm(\boldsymbol{A}_{1}+\boldsymbol{A}_{2}),\pm(\boldsymbol{A}_{1}-\boldsymbol{A}_{2})\}$.
The energy matrices $\tilde{h}_{\tilde{\alpha},\tilde{\beta}}(\tilde{\boldsymbol{\delta}})$
as well as the potential $\tilde{U}_{\alpha}(\tilde{\boldsymbol{r}})$
represented in this new basis can be constructed directly from the
original ones given in Eq.~(\ref{TB-H0}). Similar approaches of defining effective lattices (supercells) to deal with boundaries formed along chiral directions are applied to study properties of edge states along the boundaries.\cite{Farmanbar16205444}

\subsubsection{Scattering states}
\label{scatt-stat}


The reflection and the tranmission probabilities are obtained from the scattering states
as the eigenstates of the Hamiltonian for given
energies $\varepsilon$ lying in the band of the bulk part. We denote
the eigenstate of $H$ by $\left\vert \Psi\right\rangle $ and $\left\vert \phi_{\tilde{\alpha}}(\tilde{\boldsymbol{r}})\right\rangle $
the basis localized at $\tilde{\boldsymbol{r}}$ labeled by orbital $\tilde{\alpha}$.
The potential term that defines the scattering area obeys $\tilde{U}_{\tilde{\alpha}}(\tilde{\boldsymbol{r}})=\tilde{U}_{\tilde{\alpha}}(\tilde{\boldsymbol{r}}+\boldsymbol{A}_{2})$
such that the system is invariant along the $\boldsymbol{A}_{2}$
direction. Therefore, writing generally, $\left\vert \Psi\right\rangle =\sum_{\tilde{\boldsymbol{r}}}\sum_{\tilde{\alpha}}\bar{\psi}_{\tilde{\alpha}}(\tilde{\boldsymbol{r}})\left\vert \phi_{\tilde{\alpha}}(\tilde{\boldsymbol{r}})\right\rangle $,
the spatial dependence of the wavefunction $\bar{\psi}_{\tilde{\alpha}}(\tilde{\boldsymbol{r}})$
can be factorized into,
\begin{align}
\bar{\psi}_{\tilde{\alpha}}(\tilde{\boldsymbol{r}})=\bar{\psi}_{\tilde{\alpha}}(m_{1},m_{2})=e^{im_{2}\Phi_{2}}{\psi}_{\tilde{\alpha}}(m_{1}),\label{factorize}
\end{align}
 where we have used $\tilde{\boldsymbol{r}}=\tilde{\boldsymbol{r}}(m_{1},m_{2})=m_{1}\boldsymbol{A}_{1}+m_{2}\boldsymbol{A}_{2}$
with $m_{1}$ and $m_{2}$ the integers specifying the spatial coordinate.
The constant phase $\Phi_{2}$ will be determined later.

We devide the wavefunction into three regions, the left side (L), the central scattering region (CSR), and the right side (R), namely,
\begin{align}
{\psi}_{\tilde{\alpha}}(m)=\left\{ \begin{array}{c}
\psi_{\tilde{\alpha}}^{L}\left(m\right),\text{}m\le m_{b}^{L}\\
\psi_{\tilde{\alpha}}^{C}\left(m\right),\text{}m_{b}^{L}+1\le m\le m_{b}^{R}-1\\
\psi_{\tilde{\alpha}}^{R}\left(m\right),\text{}m\ge m_{b}^{R}
\end{array}\right.,\label{segment}
\end{align}
where $m_{b}^{L}$ and $m_{b}^{R}$ mark the left and the right boundaries
of the scattering area. For compactness, we denote $\boldsymbol{\psi}^{X}_{m}$,
for $X=L,\mbox{ }C,\mbox{}R$, as a column vector whose components
are $\psi_{\tilde{\alpha}}^{X}\left(m\right)$, through all the orbital
$\tilde{\alpha}$. The Schrödinger equation, $H\left\vert \Psi\right\rangle =\varepsilon\left\vert \Psi\right\rangle $,
for the segmented wavefunctions projected to the local basis then
reads
\begin{subequations}
\label{scheq-eff}
\begin{align}
\varepsilon\boldsymbol{\psi}^{S}_{m} & =\tilde{\boldsymbol{h}}^{0}\boldsymbol{\psi}^{S}_{m}
+\tilde{\boldsymbol{h}}^{J}\boldsymbol{\psi}^{S}_{m+1}
+\left[\tilde{\boldsymbol{h}}^{J}\right]^{\dagger}\boldsymbol{\psi}^{S}_{m-1},\label{scheq-bulk}\\
\varepsilon\boldsymbol{\psi}^{S}_{m_{b}^{S}} & =\tilde{\boldsymbol{h}}^{0}\boldsymbol{\psi}^{S}_{m_{b}^{S}}
+\left[\tilde{\boldsymbol{h}}_{S}^{J}\right]^{\dagger}\boldsymbol{\psi}^{S}_{m_{b}^{S}-\zeta_{S}}
 +\tilde{\boldsymbol{h}}_{S}^{J}\boldsymbol{\psi}^{C}_{m_{b}^{S}+\zeta_{S}},\label{scheq-bulksbd}
\end{align}
with $m\le m_{b}^{L}-1$ and $m\ge m_{b}^{R}+1$, for the left ($S=L$)
and the right ($S=R$) region and their boundaries respectively. The
wavefunctions on the CSR follow
\begin{align}
\varepsilon\boldsymbol{\psi}^{C}_{m} & =\left(\tilde{\boldsymbol{h}}^{0}+\tilde{\boldsymbol{u}}\left(m\right)\right)\boldsymbol{\psi}^{C}_{m}
+\tilde{\boldsymbol{h}}^{J}\boldsymbol{\psi}^{C}_{m+1}+\left[\tilde{\boldsymbol{h}}^{J}\right]^{\dagger}\boldsymbol{\psi}^{C}_{m-1},\label{scheq-CSA}\\
\varepsilon\boldsymbol{\psi}^{C}_{m_{b}^{S}+\zeta_{S}} & =\left(\tilde{\boldsymbol{h}}^{0}+\tilde{\boldsymbol{u}}\left(m_{b}^{S}+\zeta_{S}\right)\right)
\boldsymbol{\psi}^{C}_{m_{b}^{S}+\zeta_{S}}+\tilde{\boldsymbol{h}}_{S}^{J}\boldsymbol{\psi}^{C}_{m_{b}^{S}+2\zeta_{S}}
\nonumber\\
&+\left[\tilde{\boldsymbol{h}}_{S}^{J}\right]^{\dagger}\boldsymbol{\psi}^{S}_{m_{b}^{S}},\label{scheq-CSAbd}
\end{align}
with $m_{b}^{L}+2\le m\le m_{b}^{R}-2$. Here $\zeta_{S}=+1$ for
$S=L$ and $\zeta_{S}=-1$ for $S=R$. The equivalent on-site energy
matrix $\tilde{\boldsymbol{h}}^{0}$ and nearest-neighbour hopping
$\tilde{\boldsymbol{h}}^{J}$ are given by
\begin{equation}
\tilde{\boldsymbol{h}}^{0}=\tilde{\boldsymbol{h}}\left(0\right)+e^{i\Phi_{2}}\tilde{\boldsymbol{h}}\left(\boldsymbol{A}_{2}\right)+e^{-i\Phi_{2}}\tilde{\boldsymbol{h}}\left(-\boldsymbol{A}_{2}\right),\label{eqonsite}
\end{equation}
\begin{equation}
\tilde{\boldsymbol{h}}^{J}=\tilde{\boldsymbol{h}}\left(\boldsymbol{A}_{1}\right)+e^{i\Phi_{2}}\tilde{\boldsymbol{h}}\left(\boldsymbol{A}_{1}+\boldsymbol{A}_{2}\right)+e^{-i\Phi_{2}}\tilde{\boldsymbol{h}}\left(\boldsymbol{A}_{1}-\boldsymbol{A}_{2}\right),
\label{hJq1D}
\end{equation}
\end{subequations} with $\tilde{\boldsymbol{h}}_{L}^{J}=\tilde{\boldsymbol{h}}^{J}$
and $\tilde{\boldsymbol{h}}_{R}^{J}=\left[\tilde{\boldsymbol{h}}^{J}\right]^{\dagger}$.
The boldface $\tilde{\boldsymbol{h}}\left(\boldsymbol{r}\right)$
and $\tilde{\boldsymbol{u}}\left(m\right)$ are matrices in the orbital
basis with $\left[\tilde{\boldsymbol{h}}\left(\tilde{\boldsymbol{\delta}}\right)\right]_{\tilde{\alpha},\tilde{\beta}}=\tilde{h}_{\tilde{\alpha},\tilde{\beta}}(\tilde{\boldsymbol{\delta}})$
and $\left[\tilde{\boldsymbol{u}}\left(m\right)\right]_{\tilde{\alpha},\tilde{\beta}}=\delta_{\tilde{\alpha},\tilde{\beta}}\tilde{U}_{\tilde{\alpha}}(\tilde{\boldsymbol{r}})$
in which the integer coefficient of $\tilde{\boldsymbol{r}}$ in front
of $\boldsymbol{A}_{1}$ is $m$. The set of equations in Eq. (\ref{scheq-eff})
is expected from the Schr\"{o}dinger equation for solving a tight-binding
model for a quasi-one-dimensional system with on-site energy and hopping
matrices given by $\tilde{\boldsymbol{h}}^{0}$ and $\tilde{\boldsymbol{h}}^{J}$
respectively. Notably even when the original energy matrices $h_{\alpha,\beta}(\boldsymbol{\delta})$ have all of their elements real (such that
time-reversal symmetry is automatically satisfied), the effective quasi-one-dimensional system can have complex hopping matrix elements, as indicated by Eq.~(\ref{hJq1D}). This is similar to the effect of a gauge field that makes the hopping carry a phase.

Following the textbook convention of solving purely one-dimensional
scattering problem, we assume the wavefunctions for the bulk part
that are to the left ($S=L$) and to the right ($S=R$) of the scattering
area take the form,
\begin{subequations}
\label{planewave}
\begin{align}
\psi_{\tilde{\alpha}}^{L}\left(m\right) & =\sum_{l\in{M}_{L}}A_{l}^{L}\tilde{y}_{\tilde{\alpha}}^{L,l}e^{im\boldsymbol{k}_{l}^{L}\cdot\boldsymbol{A}_{1}},\label{planewaveL}\\
\psi_{\tilde{\alpha}}^{R}\left(m\right) & =\sum_{l\in{M}_{R}}A_{l}^{R}\tilde{y}_{\tilde{\alpha}}^{R,l}e^{im\boldsymbol{k}_{l}^{R}\cdot\boldsymbol{A}_{1}}.\label{planewaveR}
\end{align}
\end{subequations}Here the notation ${M}_{S}$ with $S=L,$ $R$
denotes the set of the delocalized basis. In a purely one-dimensional
problem, ${M}_{S}$ consists of only two delocalized modes, the familiar
right-going and the left-going plane waves. As will see, evanescent
modes emerge from orbital multiplicity. The wave vectors $\boldsymbol{k}_{l}^{L/R}$,
the amplitudes $A_{l}^{L/R}$ and the coefficients $\tilde{y}_{\tilde{\alpha}}^{L/R,l}$ in Eq.~(\ref{planewave})
will be found by solving the Schrödinger equation using the above
ansatz.

Substituting Eqs. (\ref{planewave}) into Eq. (\ref{scheq-bulk}) yields
\begin{subequations}
\label{eigen-tilde}
\begin{equation}
\tilde{\boldsymbol{\mathcal{H}}}\left(\boldsymbol{k}_{l}^{S}\right)\tilde{\boldsymbol{y}}^{S,l}=\varepsilon\tilde{\boldsymbol{y}}^{S,l},\label{eigen-isovector}
\end{equation}
where $\tilde{\boldsymbol{y}}^{S,l}$ is a column vector of components
$\tilde{y}_{\tilde{\alpha}}^{S,l}$ with
\begin{equation}
\tilde{\boldsymbol{\mathcal{H}}}\left(\boldsymbol{k}_{l}^{S}\right)=\tilde{\boldsymbol{h}}^{0}+e^{i\boldsymbol{k}_{l}^{S}\cdot\boldsymbol{A}_{1}}\tilde{\boldsymbol{h}}^{J}+e^{-i\boldsymbol{k}_{l}^{S}\cdot\boldsymbol{A}_{1}}\left[\tilde{\boldsymbol{h}}^{J}\right]^{\dagger}.\label{eigen-Hamiltonian}
\end{equation}
\end{subequations} Here Eq. (\ref{eigen-isovector}) is the usual
eigen-equation for getting the dispersion relation for a quasi-one-dimesional
crystal whose longitudinal direction is along $\boldsymbol{A}_{1}$ with lattice constant $\left\vert\boldsymbol{A}_{1}\right\vert$.
By restoring $\Phi_{2}=\boldsymbol{k}_{l}^{S}\cdot\boldsymbol{A}_{2}$
into Eq. (\ref{eigen-Hamiltonian}) leading to
$\tilde{\boldsymbol{\mathcal{H}}}\left(\boldsymbol{k}_{l}^{S}\right)=
\sum_{\tilde{\boldsymbol{\delta}}\in\tilde{D}}
e^{i\boldsymbol{k}_{l}^{S}\cdot\tilde{\boldsymbol{\delta}}}
\tilde{\boldsymbol{h}}\left(\tilde{\boldsymbol{\delta}}\right)$, Eq. (\ref{eigen-isovector}) becomes the standard equation
for obtaining the dispersion relation for two-dimensional crystals of lattice vectors $\boldsymbol{A}_{1}$ and $\boldsymbol{A}_{2}$. The involved modes of the scattering
problem can be found from rearranging Eq. (\ref{eigen-tilde}) to
\begin{equation}
\det\left[\varepsilon\boldsymbol{1}-\tilde{\boldsymbol{\mathcal{H}}}\left(\boldsymbol{k}_{l}^{S}\right)\right]=0,\label{eigen-det}
\end{equation} where $\boldsymbol{1}$ is the identity matrix in the space of the effective orbitals.
The determinant in Eq.~(\ref{eigen-det}) results in a polynomial in $e^{i\boldsymbol{k}_{l}^{S}\cdot\boldsymbol{A}_{1}}$
of degree $2N$, where $N$ is the number of orbitals.

The solution to Eq. (\ref{eigen-det}) contains $N$ right-going modes
that include those decaying to the right and those propagating to the right
and similarly there are $N$ left-going modes. The propagating modes
correspond to real wave vector $\boldsymbol{k}_{l}^{S}$, while the evanescent
modes give a nonzero imaginary part of $\boldsymbol{k}_{l}^{S}$ whose
sign determines to which direction (the left or the right) the wave decays. For a
purely one-dimensional problem, with $N=1$, Eq. (\ref{eigen-det})
dictates that no evanescent mode exists. The result Eq. (\ref{eigen-det})
is equivalent to the key formula, Eq. (2.12) in Ref. {[}\onlinecite{Ando918017}{]}
and Eq. (4) in Ref.{[}\onlinecite{Khomyakov05035450}{]}, obtained
from relating the wavefunctions on neighbouring sites. Here we show
that Eq. (\ref{eigen-det}) can be directly arrived from the eigen-equation
for getting the dispersion relation.

The eigenvectors $\tilde{\boldsymbol{y}}^{S,l}$
are the pseudo-spins for the mode of wave vector $\boldsymbol{k}_{l}^{S}$
at the energy $\varepsilon$. If one fixes $\boldsymbol{k}_{l}^{S}$
in Eq. (\ref{eigen-tilde}), the resulting different pseudo-spin vectors
are eigenvectors of different eigenenergies and they form an orthogonal
set. In the present case, instead, it is the energy $\varepsilon$
that is fixed and the wave vectors (for propagating modes) are searched
through the iso-energy contours provided by the dispersion relation.
The non-orthogonality among $\tilde{\boldsymbol{y}}^{S,l}$ for different
$l$ is naturally anticipated. Therefore, once the incident energy
is fixed and an incident wave vector is picked up, the value of $\Phi_{2}=\boldsymbol{k}_{l}^{S}\cdot\boldsymbol{A}_{2}$
is then automatically determined and all the input required by Eq.
(\ref{eigen-det}) are fixed. The two-dimensional nature of the momenta
can then be straightforwardly handled by Eq.
(\ref{eigen-det}) in combination of the dispersion relation deducible from Eq.~(\ref{eigen-tilde}).

From a scattering point of view, the wavefunctions on both sides of
the scattering area actually contain propagating modes moving to both
directions. The distinction lies in that only the right-going (left-going)
propagating modes qualify to be in-coming modes for incident from
the left (right). Therefore, the set of modes in $M_{S}$, obtained
by solving Eq. (\ref{eigen-det}), are categorized into $M_{S}^{\text{in}}$
and $M_{S}^{\text{out}}$. For $S=L$ ($S=R$), $M_{S}^{\text{in}}$
contains only the right (left)-going propagating modes and $M_{S}^{\text{out}}$
contains all the left (right)-going modes, both propagating and evanescent.

After establishing the properties of the involved momenta $\boldsymbol{k}_{l}^{L/R}$ and pseudo-spins $\tilde{y}_{\tilde{\alpha}}^{L/R,l}$, one can proceed to find the amplitudes $A_{l}^{L/R}$. One can eliminate $\boldsymbol{\psi}^{C}_{m}$ from Eqs.
(\ref{scheq-CSA}) and (\ref{scheq-CSAbd}), leading to
\begin{subequations}
\label{wvfn-C_GF}
\begin{equation}
\boldsymbol{\psi}^{C}_{m}=\sum_{S=L,R}\tilde{\boldsymbol{G}}\left(m,m_{b}^{S}+\zeta_{S}\right)
\left[\tilde{\boldsymbol{h}}_{S}^{J}\right]^{\dagger}\boldsymbol{\psi}^{S}_{m_{b}^{S}},\label{wvfnc-C-el}
\end{equation}
 where the Green function of the CSR is
\begin{equation}
\tilde{\boldsymbol{G}}=\left[\varepsilon\boldsymbol{1}_{C}-\tilde{\boldsymbol{H}}^{C}\right]^{-1},\label{GF-def}
\end{equation}
in which
\begin{align}
\tilde{\boldsymbol{H}}_{\tilde{\alpha},\tilde{\beta}}^{C}\left(m,m'\right)&=\delta_{m,m'}\left(\tilde{\boldsymbol{h}}^{0}
+\tilde{\boldsymbol{u}}\left(m\right)\right)
\nonumber\\&
+\delta_{m+1,m'}\tilde{\boldsymbol{h}}^{J}+\delta_{m-1,m'}\left[\tilde{\boldsymbol{h}}^{J}\right]^{\dagger},\label{GF-CSA-H}
\end{align}
\end{subequations} and $\boldsymbol{1}_{C}$ is the identity matrix in both effective site and orbital space for the CSR.
Inserting Eq.~(\ref{wvfn-C_GF}) to Eq.~(\ref{scheq-bulksbd}), which are then
substituted by the ansatz Eq.~(\ref{planewave}), we are left with equations for
the amplitudes $A_{l}^{L/R}$. Defining
\begin{equation}
\bar{A}_{l}^{S}=e^{i\boldsymbol{k}_{l}^{S}\cdot(m_{b}^{S}+\zeta_{S})\boldsymbol{A}_{1}}A_{l}^{S},\label{phase-1}
\end{equation}
we arrive at
\begin{equation}
\left(\begin{array}{cc}
\bar{\boldsymbol{W}}_{LL}^{\text{out}} & \bar{\boldsymbol{W}}_{LR}^{\text{out}}\\
\bar{\boldsymbol{W}}_{RL}^{\text{out}} & \bar{\boldsymbol{W}}_{RR}^{\text{out}}
\end{array}\right)\!\!\left(\!\!\begin{array}{c}
\bar{\boldsymbol{A}}^{L,\text{out}}\\
\bar{\boldsymbol{A}}^{R,\text{out}}
\end{array}\!\!\right)\!\!=-\!\!\left(\!\!\begin{array}{cc}
\bar{\boldsymbol{W}}_{LL}^{\text{in}} & \bar{\boldsymbol{W}}_{LR}^{\text{in}}\\
\bar{\boldsymbol{W}}_{RL}^{\text{in}} & \bar{\boldsymbol{W}}_{RR}^{\text{in}}
\end{array}\!\!\right)\!\!\left(\!\!\begin{array}{c}
\bar{\boldsymbol{A}}^{L,\text{in}}\\
\bar{\boldsymbol{A}}^{R,\text{\text{in}}}
\end{array}\!\!\right),\label{W-matrix-eq}
\end{equation}
where $\bar{\boldsymbol{A}}^{S,\text{out}/\text{in}}$ is a column
vector of components $\bar{A}_{l}^{S}$ with $l\in$$M_{S}^{\text{out}/\text{in}}$
for $S=L,\mbox{ }R$. The column vector $\bar{\boldsymbol{A}}^{S,\text{out}}$
is of dimension $N$ while $\bar{\boldsymbol{A}}^{S,\text{in}}$
is of dimension $N_{p}$, the number of right-going (left-going) propagating
modes. The matrices $\bar{\boldsymbol{W}}_{SS'}^{\text{out}}$ are
$N\times N$ and $\bar{\boldsymbol{W}}_{SS'}^{\text{in}}$ are $N\times N_{p}$
with $S,S'\in\left\{ L,R\right\} $. They are explicitly given by
\begin{align}
\left[\bar{\boldsymbol{W}}_{SS'}^{\text{out}/\text{in}}\right]_{\tilde{\alpha},l}&=
\Bigg(\delta_{SS'}\tilde{\boldsymbol{h}}_{S}^{J}\tilde{\boldsymbol{y}}^{S,l}
\nonumber\\&
-e^{-i\zeta_{S'}\boldsymbol{k}_{l}^{S'}\cdot\boldsymbol{A}_{1}}\tilde{\boldsymbol{h}}_{S}^{J}\tilde{\boldsymbol{G}}\left(S,S'\right)\left[\tilde{\boldsymbol{h}}_{S'}^{J}\right]^{\dagger}
\tilde{\boldsymbol{y}}^{S',l}\Bigg)_{\tilde{\alpha}},\label{W-matrix-def}
\end{align}
for $\tilde{\alpha}$ enumerating the $N$ orbitals while the mode index
$l$ in Eq. (\ref{W-matrix-def}) is taken from $l\in M_{S'}^{\text{out}/\text{in}}$.
The notation $\left(\cdot\right)_{\tilde{\alpha}}$ refers to the
$\tilde{\alpha}$ component of the column vector obtained from the operations inside
of the parenthesis. The surface Green function of the CSR is abbreviated
as $\left[\tilde{\boldsymbol{G}}\left(S,S'\right)\right]_{\tilde{\alpha},\tilde{\beta}}=\left[\tilde{\boldsymbol{G}}\right]_{\tilde{\alpha},\tilde{\beta}}\left(m_{b}^{S}+\zeta_{S},m_{b}^{S'}+\zeta_{S'}\right)$
from Eq. (\ref{GF-def}). The amplitudes for the out-going part of
the wavefunction $\bar{\boldsymbol{A}}^{S,\text{out}}$ can thus be
obtained from the amplitudes of the in-coming part $\bar{\boldsymbol{A}}^{S,\text{in}}$
through solving Eq. (\ref{W-matrix-eq}). This in turn fixes the coefficients
in Eq. (\ref{planewave}). With also the aid from Eq. (\ref{wvfn-C_GF}),
the whole wavefunction in Eq. (\ref{segment}) is specified.

The above derivation though is for the case that the scattering area
is defined by external potential that shifts the on-site energies
of the lattice sites in a certain area, one can straightforwardly
generalize it to the following situation: the composition of the orbitals
as well as the hopping between neighbouring sites are also different
from the rest part of the two-dimensional system. Such a generalization
is done by replacing $\left(\tilde{\boldsymbol{h}}^{0}+\tilde{\boldsymbol{u}}\left(m\right)\right)$
and $\tilde{\boldsymbol{h}}^{J}$ in Eq. (\ref{scheq-CSA}) by $\tilde{\boldsymbol{h}}^{0,C}\left(m\right)$
and $\tilde{\boldsymbol{h}}^{J,C}\left(m\right)$ respectively, in
which $\tilde{\boldsymbol{h}}^{0,C}\left(m\right)$ represent the
on-site energy matrix for the special set of orbitals on site $m$
and $\tilde{\boldsymbol{h}}^{J,C}\left(m\right)$ the hopping from
site $m$ to $m+1$ in this region. Correspondingly, in Eq. (\ref{scheq-CSAbd}),
the hopping $\tilde{\boldsymbol{h}}_{S}^{J}$ in front of $\boldsymbol{\psi}^{C}_{m_{b}^{S}+2\zeta_{S}}$
is to be replaced by $\tilde{\boldsymbol{h}}^{J,C}\left(m_{b}^{L}+1\right)$
for $S=L$ and by $\left[\tilde{\boldsymbol{h}}^{J,C}\left(m_{b}^{R}-1\right)\right]^{\dagger}$
for $S=R$. The hopping $\left[\tilde{\boldsymbol{h}}_{S}^{J}\right]^{\dagger}$
in front of $\boldsymbol{\psi}^{S}_{m_{b}^{S}}$ in Eq.
(\ref{scheq-CSAbd}) as well as its hermitian conjugate in front of
$\boldsymbol{\psi}^{C}_{m_{b}^{S}+\zeta_{S}}$ in Eq. (\ref{scheq-bulksbd})
is to be replaced by the new hopping that describes the junction between
these two regions of properly different atomic structures. These
situations can be realized by applying regional strains, formations
of grain boundaries, or regional substitutions by foreign elements.

\subsubsection{Current conservation and probabilities of reflections and transmissions}
\label{crnt-ref-trsm}

Being the eigenstate of $H$, the system prepared at $\left\vert \Psi\right\rangle $
will only evolve trivially to time $t$ by $e^{-iHt}\left\vert \Psi\right\rangle =e^{-i\varepsilon t}\left\vert \Psi\right\rangle $.
The expectation value of any observable will just remain time-independent
under such circumstance. Therefore, the charge and the current distribution
in the state $\left\vert \Psi\right\rangle $ are stationary. Below
we show how the reflection and transmission probabilities can be derived
from this property.

The charge operator, $Q_{\text{en}}$, for the charge occupation enclosed
in certain area of the system is defined by,
\begin{equation}
Q_{\text{en}}=\sum_{\tilde{\boldsymbol{r}}\in B}\sum_{\tilde{\alpha}}\left\vert \phi_{\tilde{\alpha}}(\tilde{\boldsymbol{r}})\right\rangle \left\langle \phi_{\tilde{\alpha}}(\tilde{\boldsymbol{r}})\right\vert .\label{en-crnt-opt}
\end{equation}
In Eq. (\ref{en-crnt-opt}), $B$ denotes the area bounded in the
$\boldsymbol{A}_{1}$ direction betwen the positions $\tilde{\boldsymbol{r}}(n_{L}^{0},m_{2})$
and $\tilde{\boldsymbol{r}}(n_{R}^{0},m_{2})$ with no bounds on $m_{2}$.
The two integers $n_{L}^{0}$ and $n_{R}^{0}$ mark the bounds. From
the Heisenberg equation, the current operator obtained by its definition
as the time changing-rate of charge, reads,
\begin{equation}
I_{\text{en}}=-i\left[Q_{\text{en}},H\right].\label{en-crnt-opt-1}
\end{equation}
The expectation value of the current flowing out of $B$ at time $t$
evaluated on a certain state $\left\vert \psi\right\rangle $ is given
by $\frac{d}{dt}\left[\left\langle \psi\right\vert e^{iHt}Q_{\text{en}}e^{-iHt}\left\vert \psi\right\rangle \right]=\left[\left\langle \psi\right\vert e^{iHt}I_{\text{en}}e^{-iHt}\left\vert \psi\right\rangle \right]$.
If $\left\vert \psi\right\rangle =\left\vert \Psi\right\rangle $
as an eigenstate of $H$, then $\left\langle \Psi\right\vert e^{iHt}Q_{\text{en}}e^{-iHt}\left\vert \Psi\right\rangle $
becomes time-indenpent, leading to
\begin{equation}
\left\langle \Psi\right\vert I_{\text{en}}\left\vert \Psi\right\rangle =0.\label{crntconserv1}
\end{equation}

By Eq.~(\ref{TB-H0-eff}), we find from Eq. (\ref{en-crnt-opt}) and Eq.
(\ref{en-crnt-opt-1}) that
\begin{equation}
I_{\text{en}}=I_{\text{en}}^{L}+I_{\text{en}}^{R},\label{crnt-side-sum}
\end{equation}
where
\begin{widetext}
\begin{align}
I_{\text{en}}^{S}
=\zeta_{S}\sum_{\tilde{\alpha},\tilde{\beta}}\sum_{m_{2}}\sum_{n=\left\{ 0,\pm1\right\} }
\Big\{
i\left\vert \phi_{\tilde{\alpha}}(n_{S}^{0},m_{2})\right\rangle \tilde{h}_{\tilde{\alpha},\tilde{\beta}}\left(\zeta_{S}\boldsymbol{A}_{1}+n\boldsymbol{A}_{2}\right)\left\langle \phi_{\tilde{\beta}}(n_{S}^{0}+\zeta_{S},m_{2}+n))\right\vert
 +\text{h.c.}\Big\},\label{crnt-side}
\end{align}
\end{widetext}
for $S=L$, $R$ in which we have explicitly written the integers
for the position vectors of the local basis. The operators, Eq. (\ref{crnt-side}),
depend only on the hopping from the bounds marked by $n_{L}^{0}$
and $n_{R}^{0}$ to sites one step interior to the enclosed area.
By choosing the area $B$ wide enough such that the bounds of the
CSR are within the bounds marked by $n_{L}^{0}$ and $n_{R}^{0}$,
then the wavefunctions on the CSR do not contribute to the expectation
value $\left\langle \Psi\right\vert I_{\text{en}}\left\vert \Psi\right\rangle $.
If further $n_{L}^{0}$ and $n_{R}^{0}$ are separated from $m_{b}^{L}$
and $m_{b}^{R}$ far enough such that the contributions from the evanescent
modes to $\left\langle \Psi\right\vert I_{\text{en}}\left\vert \Psi\right\rangle $
can be ignored, then $\left\langle \Psi\right\vert I_{\text{en}}\left\vert \Psi\right\rangle $
will become independent of $n_{L}^{0}$ and $n_{R}^{0}$. Explicitly
using the scattering state $\left\vert \Psi\right\rangle $ discussed
in Sec.~\ref{scatt-stat} with Eq. (\ref{crnt-side}), we obtain
\begin{equation}
\left\langle \Psi\right\vert I_{\text{en}}^{S}\left\vert \Psi\right\rangle =\zeta_{S}\sum_{m_{2}}\sum_{l\in M_{p}}\left\vert \bar{A}_{l}^{S}\right\vert ^{2}\mathcal{I}_{l},\label{crnt-side-evl}
\end{equation}
where $M_{p}$ contains all the propagating modes, both right-going
and left-going ones and
\begin{equation}
\mathcal{I}_{l}=-2\text{Im}\left\{ e^{i\boldsymbol{k}_{l}\cdot\boldsymbol{A}_{1}}\left[\tilde{\boldsymbol{y}}\left(\boldsymbol{k}_{l}\right)\right]^{\dagger}\tilde{\boldsymbol{h}}^{J}\tilde{\boldsymbol{y}}\left(\boldsymbol{k}_{l}\right)\right\} ,\label{Blochvel}
\end{equation}
is just the Bloch velocity. In Eq. (\ref{Blochvel}), we have ignored
the $S$ superscript for designating the left or the right, $\boldsymbol{k}_{l}^{S}\rightarrow\boldsymbol{k}_{l}$
and $\tilde{\boldsymbol{y}}^{S,l}\rightarrow\tilde{\boldsymbol{y}}\left(\boldsymbol{k}_{l}\right)$,
since $\boldsymbol{k}_{l}^{L}$ and $\boldsymbol{k}_{l}^{R}$ are
the same for a given propagating mode $l$. Using Eq. (\ref{crnt-side-evl})
with Eq. (\ref{crntconserv1}), we arrive at,
\begin{equation}
\sum_{l^{\mp}\in M_{\mp}}\mathcal{R}_{l_{0}^{\pm}\rightarrow l^{\mp}}+\sum_{l\in M_{\pm}}\mathcal{T}_{l_{0}^{\pm}\rightarrow l^{\pm}}=1,\label{crntconserv2}
\end{equation}
upon setting $\left\vert \bar{A}_{l_{0}^{\pm}}^{S^{\pm}}\right\vert ^{2}=1$
for incident from side $S^{\pm}$ ( with $S^{+}=L$ and $S^{-}=R$)
at the mode $l_{0}^{\pm}$ ( where the superscript + stands for right-going
and - for left-going), in which
\begin{equation}
\mathcal{R}_{l_{0}^{\pm}\rightarrow l^{\mp}}=\left\vert \bar{A}_{l^{\mp}}^{S^{\pm}}\right\vert ^{2}\frac{\mp\mathcal{I}_{l^{\mp}}}{\pm\mathcal{I}_{l_{0}^{\pm}}},\mathcal{T}_{l_{0}^{\pm}\rightarrow l^{\pm}}=\left\vert \bar{A}_{l^{\pm}}^{S^{\pm}}\right\vert ^{2}\frac{\pm\mathcal{I}_{l^{\pm}}}{\pm\mathcal{I}_{l_{0}^{\pm}}},\label{ref-trsm-basic}
\end{equation}
are the reflection and the transmission probabilities respectively.
Note that the Bloch velocities of right-going modes are positive,
namely, $\mathcal{I}_{l^{+}}>0$ while that of left-going modes are
negative, $-\mathcal{I}_{l^{-}}>0$. Therefore, if the amplitudes,
$\bar{A}_{l}^{L/R}$, obtained from Eq. (\ref{W-matrix-eq}), specify
an eigenstate of $H$, then substituting them into Eq. (\ref{ref-trsm-basic})
shall lead to the fulfillment of Eq. (\ref{crntconserv2}). The
current conservation stated in form of Eq. (\ref{crntconserv2})
can thus be used to justify the validity of the calculations, which
has been ensured in all our numerical calculations.

\subsection{Specification of computational setups }

\label{appxsec_spf}

\subsubsection{computation of incident and out-going angles}

Due to the possible warping of the dispersion relation,
the direction of the momenta on the fermi contour may not be aligned
with the direction of actual motion of the electron in that momentum
states. To unambiguously define the out-going angle
as well as the incident angle, we need to compute the velocity expectation
value for a given momentum.

In the above discussions, in order to describe scattering with a potential
that has a chiral orientation, we have introduced the effective
lattice. The in-coming states as well as the out-going states are well-defined
momentum states of the original crystal system. Therefore, to
find their velocity expectation values, we have to restore to the
original two-dimensional crystal. The dispersion relation is then
obtained from diagonalizing, $\mathcal{H}\left(\boldsymbol{k}\right)$,
namely, \begin{subequations}
\label{eigen-notilde}
\begin{equation}
\mathcal{H}\left(\boldsymbol{k}\right)\boldsymbol{y}\left(\boldsymbol{k}\right)=\varepsilon\boldsymbol{y}\left(\boldsymbol{k}\right),\label{eigen-isovector-1}
\end{equation}
where
\begin{equation}
\mathcal{H}\left(\boldsymbol{k}\right)=\sum_{\boldsymbol{\delta}\in{D}}e^{i\boldsymbol{k}\cdot\boldsymbol{\delta}}\boldsymbol{h}\left(\boldsymbol{\delta}\right).\label{eigen-Hamiltonian-1}
\end{equation}
\end{subequations} in which the matrix elements $[\boldsymbol{h}\left(\boldsymbol{\delta}\right)]_{\alpha,\beta}=h_{\alpha,\beta}$ and $D$ are those in Eq. (\ref{TB-H0}). Here $\boldsymbol{y}\left(\boldsymbol{k}\right)$
is the pseudo-spin at moment $\boldsymbol{k}$ with component $y_{\alpha}\left(\boldsymbol{k}\right)$
at orbital $\alpha$.\cite{note3}

Since here we consider only intra-band elastic scattering, the velocity
expectation value $\boldsymbol{v}\left(\boldsymbol{k}\right)$ for
$\boldsymbol{k}$ of the involved band is just the diagonal elment
indexed by the band of the usual velocity matrix\cite{Johnson732275,Smith795019,Voon9315500} (not to be confused
with the quasi-one-dimensional Bloch velocity, Eq.~(\ref{Blochvel}), previously discussed
), namely,
\begin{equation}
\boldsymbol{v}\left(\boldsymbol{k}\right)=
\left[\boldsymbol{y}\left(\boldsymbol{k}\right)\right]^{\dagger}
\left[\boldsymbol{\nabla}_{\boldsymbol{k}}\mathcal{H}\left(\boldsymbol{k}\right)\right]\boldsymbol{y}\left(\boldsymbol{k}\right).
\label{velocity-2D}
\end{equation}
The two-dimensional vector $\boldsymbol{v}\left(\boldsymbol{k}\right)$
can thus be used to compute the angle $\theta$ with respect to the normal of the interface for a given momentum
$\boldsymbol{k}$, in one-to-one correspondence.

\subsubsection{Characterisation of the potential structure}

For clarity, we concentrate on the potential profiles that are step-like and we consider $U_{\alpha}(\boldsymbol{r})=U(\boldsymbol{r})$ for all $\alpha$.
Explicitly, we describe such a potential structure
by the following (see also profiles in Fig. 1 of the main text),
\begin{align}
 & U(\boldsymbol{r})=\times\nonumber \\
 & \left\{ \begin{array}{cc}
\frac{U_{g}}{2}\left[1+\tanh\left(\frac{\ln(2/\kappa-1)}{2S_{}}(\boldsymbol{r}\cdot\boldsymbol{n}-x_{cL})\right)\right], & \text{for}~\boldsymbol{r}\in C_{L}\\
U_{g}, & \text{for}~\boldsymbol{r}\in C_{0}\\
\frac{U_{g}}{2}\left[1+\tanh\left(-\frac{\ln(2/\kappa-1)}{2S_{}}(\boldsymbol{r}\cdot\boldsymbol{n}-x_{cR})\right)\right], & \text{for}~\boldsymbol{r}\in C_{R}
\end{array}\right.,
\end{align}
 where $\boldsymbol{n}=\boldsymbol{A}_{1}/\left\vert \boldsymbol{A}_{1}\right\vert$ and the regions separating the profiles of the potential
along this lateral direction are $C_{L}=[x_{bL},x_{bL}+S_{}]$, $C_{0}=(x_{bL}+S_{},x_{bL}+W]$
and $C_{R}=(x_{bL}+W,x_{bL}+W+S_{}]$, in which $x_{bL}$
is the $\boldsymbol{A}_{1}$ coordinate of the left boundary of the scattering area. The
coordinate of the center of $C_{L}$ is $x_{cL}=x_{bL}+S_{}/2$ and
that of $C_{R}$ is $x_{cR}=x_{bL}+W+S_{}/2$. The parameter $\kappa$
controls how fast the profile in $C_{L}$ and $C_{R}$ raises (lowers)
to $U_{g}$. Here we set $\kappa=0.01$ such
that from the edge of $C_{L/R}$ it takes around a length of $S_{}$
to reach $U_{g}$. The potential $U(\boldsymbol{r})$ is zero elsewhere
when $\boldsymbol{r}$ is not in any of the regions $C_{L}$, $C_{0}$,
and $C_{R}$.




\begin{thebibliography}{References}
\bibitem{note3}The dispersion relation of the
original crystal described by Eq. (\ref{eigen-notilde}) is well embedded
in Eq. (\ref{eigen-tilde}) for the effective one. The pseudo-spins
are connected by $\tilde{y}_{\tilde{\alpha}}^{l}=e^{i\boldsymbol{k}_{l}\cdot\boldsymbol{\Delta}}y_{\alpha}\left(\boldsymbol{k}_{l}\right)$,
where $\boldsymbol{\Delta}$ points from the internal origin of
an effective lattice site to the original lattice site specified by
$\tilde{\alpha}$ whose corresponding orbital is $\alpha$.
\end{thebibliography}
\end{document}